\documentclass{JHEP3}
\title{Divergence Cancellation and Loop  Corrections in String Field Theory on a Plane Wave Background}

\author{Gianluca Grignani, Marta Orselli\\Dipartimento di Fisica and Sezione
I.N.F.N., Universit\`a di Perugia, Via A. Pascoli I-06123, Perugia,
Italia. \email{E-mail:grignani,orselli$@$pg.infn.it}
\thanks{Work supported in part by INFN and MURST of Italy.}}

\author{Bojan Ramadanovic, Gordon W. Semenoff and Donovan Young\\Department of Physics and Astronomy,
University of British Columbia, 6224 Agricultural Road, Vancouver,
British Columbia V6T 1Z1 Canada. \email{E-mail: bramadan,gordonws,dyoung$@$phas.ubc.ca}
\thanks{Work supported in part by NSERC of Canada and the Pacific Institute for
Theoretical Physics Collaborative Research Team on  Strings and
Particles.}}

\abstract{ We investigate the one-loop energy shift $\delta E$ to certain
two-impurity string states in light-cone string field theory on a plane wave
background.
We find that there exist logarithmic divergences in the sums
over intermediate mode numbers which cancel between the cubic
Hamiltonian and quartic ``contact term''. Analyzing the impurity
non-conserving channel we find that the non-perturbative 
${\cal O} ( g_2^2 \sqrt{\lambda'} )$ contribution to $\delta E/\mu$
predicted in \cite{Roiban:2002xr} is in fact an artifact of
these logarithmic divergences and vanishes with them, leaving
an ${\cal O} ( g_2^2 \lambda' )$ contribution. Exploiting the supersymmetry 
algebra, we present a form for the energy shift which appears to be manifestly convergent
and free of non-perturbative terms. We use this form to argue that 
$\delta E/\mu$ receives ${\cal O} ( g_2^2 \lambda' )$ contributions at every
order in intermediate state impurities.}

\keywords{String Field Theory, AdS/CFT Correspondence, pp-wave background}

\preprint{}

\begin{document}

\newcommand{\Tr}{\ensuremath{\mathop{\mathrm{Tr}}}}
\newcommand{\sign}{\ensuremath{\mathop{\mathrm{sign}}}}
\newcommand{\del}{\partial}
\def\be{\begin{equation}}
\def\ee{\end{equation}}
\def\bea{\begin{eqnarray}}
\def\eea{\end{eqnarray}}
\def\la{\langle}
\def\ra{\rangle}
\def\a{\alpha}
\def\dag{\dagger}
\def\wt{\widetilde}

\def\da{\dot{\alpha}}
\def\db{\dot{\beta}}
\def\dg{\dot{\gamma}}
\def\dd{\dot{\delta}}
\def\G{\Gamma}
\def\D{\Delta}
\def\L{\Lambda}
\def\S{\Sigma}
\def\a{\alpha}
\def\b{\beta}
\def\g{\gamma}
\def\d{\delta}
\def\e{\varepsilon}
\def\m{\mu}
\def\tr{\mbox{tr}}
\def\n{\nu}
\def\s{\sigma}
\def\r{\rho}
\def\l{\lambda}
\def\t{\tau}
\def\o{\omega}
\def\O{\Omega}
\def\v{\varrho}
\def\vt{\vartheta}
\def\mc{\mathcal}
\def\N{\nabla}
\def\p{\partial}
\def\K{\widetilde{K}}
\def\tb{\tilde{b}}
\def\lr{\Leftrightarrow}

\def\xxx#1 {{\sf hep-th/#1} }
\def\tr{\text{tr}}
\def\I{\mathcal I}
\def\G{\Gamma}
\def\D{\Delta}
\def\L{\Lambda}
\def\S{\Sigma}
\def\a{\alpha}
\def\b{\beta}
\def\g{\gamma}
\def\d{\delta}
\def\e{\varepsilon}
\def\m{\mu}
\def\n{\nu}
\def\s{\sigma}
\def\r{\rho}
\def\l{\lambda}
\def\t{\tau}
\def\o{\omega}
\def\O{\Omega}
\def\v{\varrho}
\def\k{\kappa}
\def\vt{\vartheta}
\def\mc{\mathcal}
\def\N{\nabla}
\def\p{\partial}
\def\la{\langle}
\def\ra{\rangle}
\def\dag{\dagger}
\def\wt{\widetilde}
\def\da{\dot{\a}}
\def\db{\dot{\b}}
\def\dg{\dot{\g}}
\def\dd{\dot{\d}}
\def\ds{\dot{\s}}

\section{Preamble}

The AdS/CFT correspondence asserts an exact duality between ${\cal
N}=4$ supersymmetric Yang-Mills theory and IIB superstring theory on
AdS$_5\times$S$^5$ background~\cite{Juan,GKP,EW}. Attempts to test
and exploit this strong coupling -- weak coupling duality have led
to a number of interesting insights about both string theory and
gauge theory.

The Penrose limit of AdS$_5\times$S$^5$  which produces a pp-wave
geometry~\cite{Blau:2001ne,Blau:2002dy} and the corresponding BMN
limit of super-Yang-Mills theory~\cite{Berenstein:2002jq} provide
one of the most important tests of the AdS/CFT correspondence.
Non-interacting type IIB string theory can be solved on the
pp-wave background using the light-cone gauge~\cite{Metsaev}.
Operators of Yang-Mills theory which correspond to the free string
states can be identified. The light-cone momenta of string theory
are
\begin{equation} p^-=\mu\left( \Delta-J\right) \label{pminus}
\end{equation}\begin{equation}\label{pplus} p^+=\frac{
\Delta+J }{2 \mu \sqrt{ g_{YM}^2N } \alpha' }
\end{equation} where $\Delta$ is the dilatation operator, J
is a U(1) R-charge and $\mu$ is a parameter of the pp-wave metric.
The limits $N,\Delta,J\to\infty$, must be taken in such a way that
$(p^+,p^-)$ remain finite. Eq.~(\ref{pminus}) relates eigenvalues
of the light-cone Hamiltonian to the eigenvalues of the dilatation
operator of ${\cal N}=4$ supersymmetric Yang-Mills theory.
Eq.~(\ref{pplus}), together with the relation between coupling
constants, $g_{YM}^2=4\pi g_s$, yield two effective couplings

\begin{equation}
\frac{1}{\left(\mu\alpha'p^+\right)^2}=\frac{g_{YM}^2N}{J^2}
\equiv\lambda' ~~,~~ 4\pi g_s\left(\mu\alpha'p^+\right)^2 =
\frac{J^2}{N} \equiv g_2~~,~~N,J\to\infty
\end{equation}
Here, $\lambda'$ is related to the string tension, and $g_2$
weights the genus of the string world-sheet. If $g_2$ is put to
zero, strings still propagate on the pp-wave background, but are
non-interacting. This free string limit coincides with the planar
limit, or large $N$ 't Hooft limit of Yang-Mills theory. This is
consistent with the fact that the remaining parameter $\lambda'$
depends on the $g_{YM}$ and $N$ only through the 't Hooft coupling
$g_{YM}^2N$.

There is a beautiful agreement between the spectra of free strings
propagating on the plane-wave background and the eigenvalues of
the dilatation operator in planar super-Yang-Mills theory in the
BMN
limit~\cite{Berenstein:2002jq,Gross:2002su,Santambrogio:2002sb}.
This agreement has been extended to scenarios beyond the BMN limit~\cite{Gubser:2002tv,Minahan:2002ve,
Beisert:2003xu,Bena:2003wd,Callan:2003xr} and to the non-perturbative sector~\cite{Green:2005pg,Green:2005rh}.
It has thus led to many promising insights.

Non-planar corrections in Yang-Mills theory should correspond to
string loop corrections in string theory.  In Yang-Mills theory in
the BMN limit, these were studied early on~\cite{Kristjansen:2002bb,
Constable:2002hw,Bianchi:2002rw,Beisert:2002bb,Constable:2002vq}
and predictions of string loop corrections to the spectra of some
string states were computed. For example, in a double expansion in
$\lambda'$ and $g_2$, the spectrum of a particular two impurity
(or two oscillator) state of the string is~\cite{Beisert:2002ff,Beisert:2003tq}
\begin{equation}
\Delta-J=2+n^2\lambda'-\frac{1}{4}n^4\lambda'^2+\frac{1}{8}n^6
{\lambda'}^3...+\frac{g_2^2}{4\pi^2} \left(
\frac{1}{12}+\frac{35}{32n^2\pi^2}\right)\left(\lambda'-
\frac{1}{2}{\lambda'}^2 n^2\right) +\ldots \label{ymspec}\end{equation}

There have been many attempts to reproduce these corrections
within string theory
~\cite{Spradlin:2002ar,Spradlin:2002rv,Chu:2002eu,Pankiewicz:2002gs,
 Dobashi:2002ar,Gomis:2002wi,Pankiewicz:2002tg,
Chu:2002wj,He:2002zu,Roiban:2002xr,
Gomis:2003kj,DiVecchia:2003yp,Pankiewicz:2003kj,Spradlin:2003xc,Gutjahr:2004dv,Dobashi:2004ka}.
This would constitute a highly nontrivial check of the AdS/CFT
correspondence at the level of interacting strings. Despite much
optimism, the present status of this work is that the order
$g_2^2\lambda'$ and order $g_2^2{\lambda'}^2$ terms in the
predicted spectrum (\ref{ymspec}) have not yet been computed using
string theory techniques alone.

Computations of string theory interactions on pp-wave backgrounds
necessarily involve light-cone string field theory. Due to
complications with the Ramond-Ramond background field, a conformal
field theory approach is not available. Light-cone string field
theory begins with constructing the light-cone Hamiltonian and the
supercharges of the residual supersymmetry of the light-cone
frame. The quadratic, ``free'' part of the Hamiltonian and
supercharges are straightforward to obtain.  They are simply a
summary of the known spectrum of free strings.

It is necessary to find the interaction parts of the Hamiltonian
and similarly the non-quadratic parts of the supercharges. The
guide to finding these is that they must respect the symmetries of
the background at the quantum level.  For the closed bosonic
string, this is achieved by a local three-string vertex.  In that
case, there is a further check of the correctness of the ansatz in
that it is known that the integration over string interactions
maps onto conformal field theory integrals over the moduli of
punctured Riemann surfaces with the appropriate vertex
operators inserted at the punctures~\cite{D'Hoker:1987pr}.

It has also been successful in superstring theory on the
background of Minkowski space where the supersymmetry algebra,
together with locality are sufficient to fix the interaction
Hamiltonian. In that case, the three-string vertex has complicated
pre-factors and four- and higher-point contact interactions appear
in the Hamiltonian.  One important role of the contact
interactions is to cancel divergences which occur at the
boundaries of the integration over the parameters of string
diagrams which are constructed using the cubic vertex.  Another
intuitive reason for why they should be there is the expectation
that the vacuum state of supersymmetric string field theory is
stable.

The problem of the pp-wave background is that, though there is a
beautiful and unambiguous free string theory available in the
light-cone gauge, all of the details of the interaction
Hamiltonian are not fixed by the supersymmetry algebra of the
background. An unknown pre-factor of the three-string vertex and
the contact interactions remains undetermined.

In the current state of the art, computations make use of an
un-justified truncation of the string spectrum to the impurity preserving channel,
which for a ``two-impurity'' external state amounts in keeping
intermediate states with only two impurities. It has been observed that some
multi-impurity states actually have contributions which are of lower
than the leading order in the coupling constant, like
$g_2^2\sqrt{\lambda'}$.  When the coupling $\lambda'$ is small, this
contribution is larger than the expected perturbative shift, which
should be of order $g_2^2\lambda'$.  It is argued that, since it is
not analytic in the coupling, it corresponds to a non-perturbative
correction to Yang-Mills theory.

In this Paper, we make two bits of progress toward matching
string field theory and Yang-Mills theory in the BMN limit.  First
of all, we observe that, in string field theory, individual
perturbative contributions to some processes have logarithmic
divergences when summed over intermediate states.   These
quantities appear to be finite on the Yang-Mills theory side.
Then, we note that, when all contributions are assembled, these
divergences cancel, leaving a finite result. We present an algebraic
proof of cancellation of divergences for some two-impurity states.
In fact for the spectrum of these states, the leading
order contribution of order $g_2^2\sqrt{\lambda'}$ cancels
along with the logarithmic divergences, leaving the natural leading
order of $\lambda'$.  Unfortunately, at this point we cannot go
beyond this observation.  The terms of order $\lambda'$ seem to
obtain contribution from intermediate states with any number of
impurities, making their precise computation a formidable task.

It is worth noting that in the DLCQ of type-IIB superstring on the plane-wave background, 
for which there exists a dual gauge theory~\cite{Mukhi:2002ck},
and the mass shift corrections to two
impurity operators have been computed~\cite{DeRisi:2004bc},
it was shown that energy shifts and corrections 
appear in the combination 
$(4 \pi g_s\mu \alpha' p^+ )^2=g_2^2\lambda'$~\cite{Sheikh-Jabbari:2004ik}.

We use the notation and conventions of
ref.~\cite{Gutjahr:2004dv} and reproduce them in appendices 
\ref{freestrings} through \ref{neu}.

\section{Invitation: impurity conserving channel}
\label{trace}

The shift in energy of a string state is computed using quantum
mechanical perturbation theory. The Hamiltonian has a known
quadratic part $H_2$ and interaction terms $g_2H_3$, $g_2^2H_4$,
etc.  $g_2^2H_4$ and higher order are ``contact terms''.  The
cubic interaction $H_3$ has non-zero matrix elements only between
states with $n$ strings and $n\pm 1$ strings.  For this reason,
the linear order in perturbative correction to a single string
state vanishes,
$$\delta E_n^{(1)}=\langle \phi_n^A|H_3|\phi_{n}^B\rangle =0
$$ where $|\phi_{n}^A\rangle$ and $|\phi_{n}^B\rangle$ are  one-string states
and $$H_2 |\phi_{n}^A\rangle= E^{(0)}_n|\phi_{n}^A\rangle$$ If the
states are degenerate, an ortho-normal basis is labeled by
$A$ and $B$.

The leading non-vanishing correction is of second order,
 \be \delta E_n^{(2)AB}  =g_2^2 \left(
 \langle \phi_n^A|H_3\frac{{\cal P}}{E_n^{(0)}-H_{2}} H_3|\phi_n^B\rangle
+\langle \phi_n^A|H_4|\phi_n^B\rangle\right) \label{shift} \ee
where ${\cal P}$ is a projection operator onto the orthogonal
complement of the states spanned by $|\phi_{n}^A\rangle$.  It will
also be used to enforce level matching of intermediate states.

In previous literature, it is common to consider an approximation
to (\ref{shift}) which restricts to two-oscillator intermediate
states.   This is done by replacing ${\cal P}$ in the first term
with a projection onto level-matched two oscillator, two-string
states.  In the second term, supersymmetry is used to factor $H_4$
into a product of super-charges and a similar projector is
inserted (see equation (\ref{contact})). This is the so-called ``two-impurity channel'', or
``impurity conserving channel''. It is known that, although they
are very similar in form, the correction obtained in this channel
does not match the predictions of Yang-Mills theory~\cite{Gutjahr:2004dv}.

One source of discrepancy is that the expression (\ref{shift})
could obtain contributions from other than just two-impurity
intermediate states. In fact, it was noted in
ref.~\cite{Roiban:2002xr} that the contribution of the four
impurity channel to the mass shift of the string
state\footnote{The normalization of this state is
$1+\frac{1}{2}\delta^{ij}$} $$\left. |[{\bf 9}, {\bf
1}]\ra^{(ij)}\right. =
\frac{1}{\sqrt{2}}\left(\a^{\dag\,i}_n\a^{\dag\,j}_{-n}+\a^{\dag\,j}_n\a^{\dag\,i}_{-n}
-\frac{1}{2}\d^{ij}\a^{\dag\,k}_n\a^{\dag\,k}_{-n}\right)|\a\ra
$$  appeared to
diverge if the large $\mu$ limit was taken prior to summing over
intermediate mode numbers. The authors noted that summing the
expressions at finite $\mu$ regularized the divergence but then
resulted in a $\sqrt{\lambda'}$ contribution to $\delta
E^{(2)}/\mu$. For small $\lambda'$, such a contribution is more
important than the leading contributions which arose from the
two-impurity-channel, for example, where the leading terms were of
order $\lambda'$, and it is hard to see how it could ever arise in
Yang-Mills theory.

We will show that, in fact, this $\sqrt{\lambda'}$ contribution
comes from mode-number sums which are logarithmically divergent.
The existence of logarithmic divergences is counter to the
philosophy that string field theory loop corrections should be
finite.  Upon further investigation, we shall see that, in this
four-impurity channel case, the logarithmic divergences actually
cancel.  Along with the logarithmic divergences, the contributions
of order $\sqrt{\lambda'}$ also cancel, leaving what one expects,
a leading contribution of order $\lambda'$.

The cancellation of logarithmic divergences is between
contributions from the $H_3$ vertex and the contact term.  This is
in line with the known fact that the role of the contact term is
to cancel divergences of this kind, which also arise in the
conformal field theory computation of superstring amplitudes on
Minkowski space.  There, the contact terms cancel divergent
surface terms which appear upon integration by parts in the
integrals of correlators of vertex operators over the moduli of
Riemann surfaces~\cite{Green:1987qu}.

Indeed logarithmic divergences of precisely the same nature, and a
similar cancellation mechanism, can already be seen in a much
simpler case: a careful calculation of the two-impurity channel
contribution to the mass shift of the normalized bosonic trace state
\be\left|[{\bf 1}, {\bf 1}]\ra\right. =
\frac{1}{{2}}\alpha^{i\dagger}_n \alpha^{i \dagger}_{-n} |\alpha
\rangle\label{singlet}\ee In~\cite{Gomis:2003kj}, this calculation was performed
by taking the large $\mu$ limit first, then summing over mode
numbers. That procedure found a finite result.   However, it is
not quite legitimate.  If $\mu$ is kept finite, there are
logarithmically divergent summations which must be dealt with
before the large $\mu$ limit is taken.  In the following we will
re-examine this question and observe that the logarithmically
divergent pieces would make the mass shift infinite even when
$\mu$ is finite.   Happily, we shall find that they cancel when
all terms are taken into account.

 We
calculate the following matrix element for the state $\left|[{\bf
1}, {\bf 1}]\ra\right.$
\bea &&\langle \alpha_3 |\frac{1}{2}
\alpha^i_{n} \alpha^i_{-n} \, \langle \wt{\a}_2 | \langle \wt{\a}_1
| \alpha^{K}_{p} \alpha^{L}_{-p} |H_3
\rangle = -g_2\,\frac{r\,(1-r)}{8} \left[8\left(
\frac{\omega_{n}^{(3)}}{\alpha_3} +
\frac{\omega_{p}^{(1)}}{\alpha_1} \right) {\widetilde
N}_{-n\,p}^{3\,1} {\widetilde N}_{n\,p}^{3\,1} \,\delta^{kl}\right. \cr
&&\left.+16\,\frac{\omega_{n}^{(3)}}{\alpha_3} {\widetilde
N}_{n\,n}^{3\,3} {\widetilde N}_{p\,-p}^{1\,1} \,\delta^{KL} + 16
\,\frac{\omega_{p}^{(1)}}{\alpha_1} {\widetilde N}_{n\,-n}^{3\,3}
{\widetilde N}_{p\,p}^{1\,1} \,\Pi^{KL}\right] \label{trmatel} \eea

\noindent where the index $i=1,\ldots,4$ is summed over. Note that $K,L = 1,\ldots,8$, while
$\delta^{kl}$ is non-zero only for $k=l=1,\ldots,4$. The matrix
$\Pi^{KL}$ is given by
$$
\Pi^{KL} = \mbox{diag}(1,1,1,1,-1,-1,-1,-1)
$$
When calculating the $H_3$ contribution to the mass shift it is
only the very last term in (\ref{trmatel}) which is divergent.
Singling-out its contribution, one finds\footnote{We use the following intermediate state
projector:
$${\bf 1}_B=
\int_0^1\frac{dr}{2\,r(1-r)}
\sum_{p}
\a_{p}^{\dag\,K}\,\a_{-p}^{\dag\,L}\,| \wt{\a}_1\ra\,
| \wt{\a}_2\ra\la \wt{\a}_2|\,\la \wt{\a}_1|\,\a_{-p}^L\,\a_{p}^K$$
where $\wt{\a}_1\equiv -\alpha_3 \,r$ and 
$\wt{\a}_2\equiv -\alpha_3 (1-r)$. Note that oscillators act only on the vacuum
closest to them.},

\begin{equation}\label{divH3}
\delta E^{\mbox{div}}_{H_3} =  \int_0^1 dr\,\left(g_2 \frac{r(1-r)}{8} \right)^2
\,\frac{-\alpha_3}{2\,r\,(1-r)} \sum_{KL} \sum_{p=-\infty}^{\infty}\frac{
\left[ 16\,\frac{\omega_{p}}{-r\,\alpha_3}
{\widetilde N}_{n\,-n}^{3\,3} \,{\widetilde N}_{p\,p}^{1\,1} \,\Pi^{KL} \right]^2}
{2\,\omega_{n} -2\,r^{-1}\omega_{p} }
\end{equation}

A quick inspection of the forms of the Neumann matrices (see
Appendix D) reveals that the numerator in (\ref{divH3}) goes like
a constant for large $|p|$, and thus the sum as a whole goes like
$1/|p|$ for $|p| \gg |\mu\alpha_3|$. This is a logarithmically
diverging sum. In~\cite{Gomis:2003kj} the strict large $\mu$
limit was taken for the energy denominator, leading to a
convergent $1/p^2$ behavior instead. Here we will stick with the
finite $\mu$ expressions and show that the divergence is removed
by the contact term. Note that a double fermionic impurity
intermediate state also contributes to the $H_3$ piece, however it
does not display any divergent behavior. Further, the
$\alpha^\dagger_0 |\alpha_1\rangle \alpha^\dagger_0
|\alpha_2\rangle$ intermediate state is unimportant to us for the
same reason.

The contribution from the contact term stems from the following matrix element,
\bea
\label{contactme}
&&\left( g_2 \frac{\eta}{4} \sqrt{\frac{r\,(1-r)\, \alpha'}{-2\,\alpha_3^3}} \right)^{-1}
\langle \alpha_3 | \frac{1}{2}\alpha^i_{n} \alpha^i_{-n} \, \langle \wt{\a}_2 | \langle \wt{\a}_1 |
\alpha^{K}_{p} \beta^{\Sigma_1\,\Sigma_2}_{-p}
|Q_{3\, \beta_1 {\dot \beta}_2} \rangle\cr
&&\qquad= \biggl( G_{|p|}^{(1)} \, K_{n}^{(3)} {\widetilde N}^{3\,1}_{n\,p}
+ G_{|p|}^{(1)} \, K_{-n}^{(3)} {\widetilde N}^{3\,1}_{-n\,p}
 \biggr)( \sigma^k)^{\dot{\sigma_1}}_{ \beta_1 } \delta^{{\dot \sigma_2}}_{{\dot \beta_2}}
+4\, G_{|p|}^{(1)} \, K_{-p}^{(1)} {\widetilde N}^{3\,3}_{n\,-n} (
\sigma^K)^{\Sigma}_{ \beta } \delta^{\Sigma}_{\beta}~~~~~ \eea along
with a similar element with $| Q_{3\, {\dot \beta_1} \beta_2}
\rangle$. Here $K=1,\ldots,8$ while the $\Sigma$ and $\beta$ indices are
either dotted or undotted as required by the particular SO(4)
representation indicated by $K$. 
The last term in
(\ref{contactme}) gives rise to a log-divergent sum. For large
positive  $p$, $(K_{-p}^{(1)})^2$ goes as a constant, and so the
sum is controlled by $(G_{|p|}^{(1)})^2$ which goes as $1/p$, and
hence diverges logarithmically. For $p$ negative, the sum
converges. Thus, the divergent contribution to $\delta E^{(2)}$ is
found to be\footnote{We use the following intermediate state
projector:
$${\bf 1}_F=
\int_0^1\frac{dr}{r(1-r)}
\sum_{p}
\a_{p}^{\dag\,K}\,\b_{-p}^{\dag\,\Sigma_1\,\Sigma_2}\,| \wt{\a}_1\ra\,
| \wt{\a}_2\ra\la \wt{\a}_2|\,\la \wt{\a}_1|\,\b_{-p}^{\Sigma_1\,\Sigma_2}\,\a_{p}^K$$},

\begin{equation}\label{divH4}
\delta E^{\mbox{div}}_{H_4} = 8 \int_0^1 dr\,
\left( g_2 \frac{1}{4} \sqrt{\frac{r\,(1-r)\, \alpha'}{-2\,\alpha_3^3}} \right)^2
\frac{1}{r(1-r)} \sum_{p=1}^{\infty}
 \biggl( 4\,G_{|p|}^{(1)} \,K_{-p}^{(1)} {\widetilde N}^{3\,3}_{n\,-n}
\biggr)^2
\end{equation}
The leading factor of 8 comes from the sum over $K$.
Note that two factors of 2 from the delta function (in Pauli
indices) and the (squared) Pauli matrix trace cancel the two
factors of $1/8$ coming from the two terms of the contact term,
$Q_{3\beta_1\dot\beta_2}Q_3^{\beta_1\dot\beta_2}$ and
$Q_{3\dot\beta_1\beta_2} Q_3^{\dot\beta_1\beta_2}$ (See equation
(\ref{contact})).  Again the intermediate state $\alpha^\dagger_0
|\alpha_1\rangle \beta^\dagger_0 |\alpha_2\rangle$ is unimportant
to convergence and is ignored here.

In taking the large $p$ limits of the summands in (\ref{divH3}) and (\ref{divH4}), one finds,

\begin{equation}
\delta E^{\mbox{div}}_{H_3} \sim -\frac{1}{2} \int_0^1 dr  \,\frac{g_2^2\,r(1-r)}{r\,|\alpha_3|\,\pi^2}
\left( {\widetilde N}^{3\,3}_{n\,-n} \right)^2 \, \frac{1}{|p|}
\end{equation}

\begin{equation}
\delta E^{\mbox{div}}_{H_4} \sim +\int_0^1 dr \,\frac{g_2^2\,r(1-r)}{r\,|\alpha_3|\,\pi^2}
\left( {\widetilde N}^{3\,3}_{n\,-n} \right)^2 \, \frac{1}{p}
\end{equation}

 Noting that in the $H_3$ contribution the divergence is found for both
positive and negative $p$, while in the $H_4$ contribution the divergence occurs
only for positive $p$, and hence a relative factor of 2 is induced
in the $H_3$ term, one sees that the logarithmically divergent sums cancel
identically between the $H_3$ and contact terms, leaving a
convergent sum.

This cancellation fixes the relative weight of the $H_3$ and
contact terms to that employed in~\cite{Gutjahr:2004dv}.  It
differs by a factor of $1/2$ from the weight originally given 
in~\cite{Roiban:2002xr}, where it was argued to be a reflection
symmetry factor.

The singlet state  $\left|[{\bf 1}, {\bf 1}]\ra\right.$ is more generally
constructed also in terms of fermionic oscillators, e.g. $(\beta_n^\dagger)_{\alpha_1 \alpha_2}
(\beta_n^\dagger)^{\alpha_1 \alpha_2}$. These states start to mix with the bosonic trace
state (\ref{singlet}) at loops higher than one. Therefore the two impurity channel contribution
to the mass shift of the state (\ref{singlet}), 
can be computed, up to one loop, by assembling the finite parts of all the
possible two impurity intermediate states, both in $H_3$ and in $H_4$, without bothering about 
state mixing.
With the relative coefficient between $H_3$ and $H_4$ fixed by the requirement of the
cancellation of divergences, the result reads
\be
\frac{\delta E^{(2)}_n}{\mu}=\frac{g_2^2\lambda'}{4\pi^2}\left(\frac{1}{24}+\frac{65}{64 n^2 \pi^2}\right)
\ee
This is in agreement with the order $\lambda'$ result of ref.~\cite{Gutjahr:2004dv} 
for the traceless symmetric state $\left|[{\bf 9}, {\bf 1}]\ra^{(ij)}\right.$
and in disagreement with what is expected from the gauge theory.
In the next section we will discuss a possible explanation for this disagreement
\footnote{We note that in fact it is not hard to show that the trace (but not the 
$|[{\bf 9}, {\bf 1}]\ra^{(ij)}$) state receives an
order $\lambda'$ contribution from the zero impurity channel.}.

\section{Four impurity channel}
\label{4imp}

\indent\indent We now consider the mass shift of the $\left|[{\bf
9}, {\bf 1}]\ra^{(ij)}\right.$ string state due to intermediate
states which contain four impurities.

In the explicit expression for the matrix element to be quoted
below, we shall see that the parameter $\mu\alpha_3$ occurs only
in combinations involving $\omega_p$ and there is a duality
between the large $p$ and the large $\mu\alpha_3$ limits.
Therefore, since a logarithmic divergence in the sums indicates
that the summands have as many (inverse) powers of the summation
variables as there are summation variables, this translates into
vanishing $\mu\alpha_3$ dependence for this contribution to
$\delta E^{(2)}$, leaving $\delta E^{(2)}/\mu \sim
\sqrt{\lambda'}$. It is thus seen that $\sqrt{\lambda'}$ behavior
is simply the result of log divergences, which should, if pp-wave
light-cone string field theory is to make any sense, cancel out
entirely.

We begin with the $H_3$ contribution to the mass shift. 
We consider the following intermediate state,
\begin{equation} \label{projector}
{\bf 1}_B=
\int_0^1\frac{dr}{4!\,r(1-r)}
\sum_{p_1\,p_2\,p_3\,p_4}
\a_{p_1}^{\dag\,K}\,\a_{p_2}^{\dag\,L}\, \a_{p_3}^{\dag\,M}\,\a_{p_4}^{\dag\,N}\, | \wt{\a}_1\ra\,
| \wt{\a}_2\ra\la \wt{\a}_2|\,\la \wt{\a}_1|\,\a_{p_4}^N\,\a_{p_3}^M \,\a_{p_2}^L\,\a_{p_1}^K
\end{equation}
where the sum over mode numbers is restricted by the level matching
condition  $\sum_i p_i = 0$ and $\wt{\a}_1\equiv -\alpha_3 \,r$, 
$\wt{\a}_2\equiv -\alpha_3 (1-r)$.

Although there are many possible contractions of this state with
the oscillators in $|H_3\ra $, we will only be concerned with
those which lead to log divergent sums. These are the ones where
the $\alpha^\dag $ in the prefactor of $|H_3\ra$ contracts with one of the
oscillators in ${\bf 1}_B$. We find this contribution to $\delta
E^{(2)}$ to be\footnote{The normalization $1+\frac{1}{2}\delta^{ij}$ of
the external state has been suppressed here.}, \bea &&\delta E^{\mbox{div}}_{H_3}=\int_0^1
\frac{dr}{4!\,r\,(1-r)} \left(g_2 \frac{r(1-r)}{4} \right)^2 \,
\sum_{p_2\,p_3\,p_4} \frac{-\alpha_3\,r}{2\,\omega_n\,r -
\sum_{i=1}^4\omega_{p_i}} \times \cr &&\left(2\,
\frac{\omega_{p_1} + \omega_{p_2}}{-r\,\alpha_3} \, {\widetilde
N}^{1\,1}_{-p_1\, p_2} \right)^2
\left\{8\cdot12\,\left({\widetilde N}^{3\,1}_{n\, p_3} {\widetilde
N}^{3\,1}_{-n\, p_4} \right)^2 + 6\,{\widetilde N}^{3\,1}_{n\,
p_3} {\widetilde N}^{3\,1}_{-n\, p_3} {\widetilde N}^{3\,1}_{n\,
p_4} {\widetilde N}^{3\,1}_{-n\, p_4}\right\} \label{divH34} \eea
where $p_1 = -(p_2+p_3+p_4)$. The factors of $6$ and $12$ are
combinatoric and count the number of ways equivalent contractions
can be made. The factor of $8$ comes from a sum over the spacetime
indices of ${\bf 1}_B$ and only affects squared terms. It is easy
to see that in the above, the sum over $p_2$ is log divergent. In
fact, it is the very same form as appears in (\ref{divH3}). Using
the techniques described in Appendix \ref{largemu}, one sees that
$\delta E^{\mbox{div}}_{H_3} \sim \mbox{constant}$, and therefore
$\delta E^{(2)}/\mu \sim \sqrt{\lambda'}$. There are also
contributions from intermediate states which contain two bosonic
and two fermionic impurities, however these produce convergent
sums and ${\cal O}(\lambda')$ contributions to $\delta
E^{(2)}/\mu$.

We now show that the contact term contribution stemming from the following
intermediate state,
\begin{equation}
{\bf 1}_F=
\int_0^1\frac{dr}{3!\,r(1-r)}
\sum_{p_1\,p_2\,p_3\,p_4}
\beta_{p_1}^{\dag\,a}\,\a_{p_2}^{\dag\,L}\, \a_{p_3}^{\dag\,M}\,\a_{p_4}^{\dag\,N}\, | \wt{\a}_1\ra\,
| \wt{\a}_2\ra\la \wt{\a}_2|\,\la \wt{\a}_1|\,\a_{p_4}^N\,\a_{p_3}^M \,\a_{p_2}^L\,\beta_{p_1}^a
\end{equation}
cancels the divergent piece coming from the  $H_3$ contribution, leaving an ${\cal
O}(\lambda')$ contribution to $\delta E^{(2)}/\mu$. In the above
$a$ is an SO(8) index and thus represents both dotted and undotted
indices in the language of~\cite{Gutjahr:2004dv}. The log
divergent piece comes from contractions where the $\a^\dag$ in the
prefactor of $|Q_3\ra$ is joined with one of the bosonic
oscillators in ${\bf 1}_F$. One finds,
\bea \delta E^{\mbox{div}}_{H_4}&=&\int_0^1
\frac{dr}{3!\,r\,(1-r)} \left( g_2 \frac{1}{4}
\sqrt{\frac{r\,(1-r)\, \alpha'}{-2\,\alpha_3^3}} \right)^2
\sum_{p_2\,p_3\,p_4} \left(2\, G_{p_1} K_{-p_2} \right)^2  \cr
&&\times\left\{ 8\cdot6\,\left({\widetilde N}^{3\,1}_{n\, p_3}
{\widetilde N}^{3\,1}_{-n\, p_4} \right)^2 + 3\, {\widetilde
N}^{3\,1}_{n\, p_3} {\widetilde N}^{3\,1}_{-n\, p_3} {\widetilde
N}^{3\,1}_{n\, p_4} {\widetilde N}^{3\,1}_{-n\, p_4} \right\} \eea
In the above one sees the very same pattern as was seen in section
\ref{trace}. The sum over $p_2$ is divergent on the positive side,
and cancels the divergence in (\ref{divH34}). The remaining
(convergent) expression gives an ${\cal O}(\lambda')$ contribution
to $\delta E^{(2)}/\mu$. Again, there is a non-divergent
contribution from the intermediate state with three fermionic and
one bosonic impurity which is not considered here.

The cancellation exposed here is also found for the following remaining pairs
of intermediate states,
\bea
&&{\bf 1}_B=
\int_0^1\frac{dr}{3!\,r(1-r)}
\sum_{p_1\,p_2\,p_3}
\a_{p_1}^{\dag\,K}\,\a_{p_2}^{\dag\,L}\, \a_{p_3}^{\dag\,M}\, | \wt{\a}_1\ra\,
\a_{0}^{\dag\,N}\,| \wt{\a}_2\ra\la \wt{\a}_2|\,\a_{0}^N\,\la \wt{\a}_1|
\,\a_{p_3}^M \,\a_{p_2}^L\,\a_{p_1}^K \cr
&&{\bf 1}_F=
\int_0^1\frac{dr}{2!\,r(1-r)}
\sum_{p_1\,p_2\,p_3}
\beta_{p_1}^{\dag\,a}\,\a_{p_2}^{\dag\,L}\, \a_{p_3}^{\dag\,M}\,| \wt{\a}_1\ra\,
\a_{0}^{\dag\,N}\,| \wt{\a}_2\ra\la \wt{\a}_2|\,\a_{0}^N\,\la \wt{\a}_1| \,\a_{p_3}^M \,\a_{p_2}^L\,\beta_{p_1}^a
\eea
where $\sum_{i=1}^3 p_i = 0$ and,
\bea
&&{\bf 1}_B=
\int_0^1\frac{dr}{2\cdot (2!)^2 \,r(1-r)}
\sum_{p_1\,p_2}
\a_{p_1}^{\dag\,K}\,\a_{-p_1}^{\dag\,L}\, | \wt{\a}_1\ra\,
\a_{p_2}^{\dag\,M}\,\a_{-p_2}^{\dag\,N}\, | \wt{\a}_2\ra
\la \wt{\a}_2|\,\a_{-p_2}^N\,\a_{p_2}^M \,\la \wt{\a}_1|\,\a_{-p_1}^L\,\a_{p_1}^K\cr
&&{\bf 1}_F=
\int_0^1\frac{dr}{2! \,r(1-r)}
\sum_{p_1\,p_2}
\a_{p_1}^{\dag\,K}\,\a_{-p_1}^{\dag\,L}\, | \wt{\a}_1\ra\,
\a_{p_2}^{\dag\,M}\,\beta_{-p_2}^{\dag\,a}\, | \wt{\a}_2\ra
\la \wt{\a}_2|\,\beta_{-p_2}^a\,\a_{p_2}^M \,\la \wt{\a}_1|\,\a_{-p_1}^L\,\a_{p_1}^K
\eea and so we find that the entire contribution to $\delta
E^{(2)}/\mu$ from the four impurity channel is convergent / leads
as $\lambda'$. It is not hard to generalize the above argument to
${\bf 1}_B$'s containing an arbitrary number of bosonic impurities
and no fermionic impurities. The divergent expressions cancel
against contact interactions with ${\bf 1}_F$'s containing one
fermionic and the same number (less-one) of bosonic oscillators as
${\bf 1}_B$. Adding fermionic impurities is far less trivial 
because the full forms~\cite{Pankiewicz:2003kj} of $|H_3\ra$ and $|Q_3\ra$,
given in Appendix \ref{dynamical}, must be used for the calculation\footnote{We remind the reader that
the $|[{\bf 9}, {\bf 1}]\ra^{(ij)}$ state receives
no contributions to its energy shift from the zero impurity channel.}.
In the next section, however, a more elegant argument
is presented which claims the absence of log divergences for
arbitrary impurity intermediate states.

\section{Generalizing to arbitrary impurities}

\label{arbimp} It is possible to formally manipulate the contact
term in such a way that the $H_3$ portion of the energy shift is
canceled entirely, leaving a convergent expression, which appears
devoid of any $\sqrt{\lambda'}$ contributions to $\delta
E^{(2)}/\mu$. $H_3$ arises from the anti-commutators derived from
the dynamical constraints up to order $g_2$ (cfr. Appendix
\ref{dynamical}) \bea
&&\left\{Q_{2\alpha_1\dot\alpha_2},Q_{3\beta_1\dot\beta_2}\right\}+
\left\{Q_{3\alpha_1\dot\alpha_2},Q_{2\beta_1\dot\beta_2}\right\}
=-2\epsilon_{\alpha_1\beta_1}\epsilon_{\dot\alpha_2\dot\beta_2}
H_3~,\cr
&&\left\{Q_{2\dot\alpha_1\alpha_2},Q_{3\dot\beta_1\beta_2}\right\}+
\left\{Q_{3\dot\alpha_1\alpha_2},Q_{2\dot\beta_1\beta_2}\right\}
=-2\epsilon_{\dot\alpha_1\dot\beta_1}\epsilon_{\alpha_2\beta_2}
H_3 \label{Q2Q3} \eea Analogously to order $g^2_2$ one has \bea
\left\{Q_{3\alpha_1\dot\alpha_2},Q_{3\beta_1\dot\beta_2}\right\}
+\left\{
Q_{2\alpha_1\dot\alpha_2},Q_{4\beta_1\dot\beta_2}\right\}+\left\{
Q_{4\alpha_1\dot\alpha_2},Q_{2\beta_1\dot\beta_2}\right\}=
-2\epsilon_{\alpha_1\beta_1}\epsilon_{\dot\alpha_2\dot\beta_2}
H_4 \\
\left\{Q_{3\dot\alpha_1\alpha_2},Q_{3\dot\beta_1\beta_2}\right\}+\left\{
Q_{2\dot\alpha_1\alpha_2},Q_{4\dot\beta_1\beta_2}\right\}+\left\{
Q_{4\dot\alpha_1\alpha_2},Q_{2\dot\beta_1\beta_2}\right\}
=-2\epsilon_{\dot\alpha_1\dot\beta_1}\epsilon_{\alpha_2\beta_2}
H_4 \label{Q3Q3} \eea

To get $H_3$ and $H_4$ the first of the equations in both
(\ref{Q2Q3}) and (\ref{Q3Q3}) should be multiplied by
$\epsilon^{\alpha_1\beta_1}\epsilon^{\dot\beta_2\dot\alpha_2}$ and
the second by
$\epsilon^{\dot\alpha_1\dot\beta_1}\epsilon^{\beta_2\alpha_2}$. On
the left hand sides of the equations the epsilons just raise
indices, on the right hand sides they give -4. The
anti-commutators in eq.(\ref{Q2Q3}) thus give \be
\left\{Q_{2\beta_1\dot\beta_2},Q_3^{\beta_1\dot\beta_2}\right\}=+4
H_3~~~,~~~
\left\{Q_{2\dot\beta_1\beta_2},Q_3^{\dot\beta_1\beta_2}\right\}=+4
H_3 \label{h3} \ee Those in  eq.(\ref{Q3Q3})
give the
contact Hamiltonian \bea H_4={1\over 8}
Q_{3\beta_1\dot\beta_2}Q_3^{\beta_1\dot\beta_2}+{1\over 8}
Q_{3\dot\beta_1\beta_2} Q_3^{\dot\beta_1\beta_2}+{1\over 8}
Q_{4\beta_1\dot\beta_2}Q_2^{\beta_1\dot\beta_2}+{1\over 8}
Q_{4\dot\beta_1\beta_2} Q_2^{\dot\beta_1\beta_2}\nonumber
\\+{1\over 8}
Q_{2\beta_1\dot\beta_2}Q_4^{\beta_1\dot\beta_2}+{1\over 8}
Q_{2\dot\beta_1\beta_2} Q_4^{\dot\beta_1\beta_2} \label{H4} \eea

Using these formulas, the contribution of $H_4$ to $\delta
E^{(2)}$ can be rewritten as a sum of a term which cancels the
$H_3$ contribution plus other pieces which all contain $Q_2$
acting on one of the external states. Taking the expectation value
of part of (\ref{H4}), and introducing a representation of unity,
we have,
\bea\label{okok} {1\over 8} \left<
Q_{3\beta_1\dot\beta_2} Q_3^{\beta_1\dot\beta_2} +
Q_{3\dot\beta_1\beta_2} Q_3^{\dot\beta_1\beta_2}
\right>={1\over 8}
\left< Q_{3\beta_1\dot\beta_2}P
\frac{E_0-H_2}{E_0-H_2}Q_3^{\beta_1\dot\beta_2}\right> \\
+{1\over 8}
\left<
Q_{3\dot\beta_1\beta_2}P\frac{E_0-H_2}{E_0-H_2}Q_3^{\dot\beta_1\beta_2}\right>
\eea 
It could be that the energy denominator which we have
introduced here will have a zero.  In that case, the projector $P$ is
a reminder to define the singularity using a principle value
prescription~\footnote{  There is one additional subtlety, the
intermediate states must each obey the level-matching condition.
This condition can be enforced by inserting a projection operator.
For example, for two-string intermediate states, we can combine
such a projector with the energy denominator as
\begin{equation}
\frac{P}{E_0-H_2} = \int_0^\infty d\tau ~e^{E_0\tau}\int_{-\pi}^\pi
\frac{ d\theta_1}{2\pi} \int_{-\pi}^\pi \frac{d\theta_2}{2\pi}
~e^{-H_2^{(1)}\tau+i \theta_1 N^{(1)}}~e^{-H_2^{(2)}\tau+i\theta_2
N^{(2)}}
\end{equation}
where \begin{equation} N^{(r)}=\sum_n n\left(
a^{I(r)\dagger}_n a^{I(r)}_n+b^{(r)\dagger}_{an}b^{(r)}_{an}\right)
\end{equation}
with $r=1,2$ are the level number operators for the two
intermediate strings.  The net effect of the operators in the
above equation is to make the replacement
$\left(a_n^{(r)\dagger},b_n^{(r)\dagger}\right)\to\left(
e^{-\omega_n\tau+in\theta_{(r)}} a_n^{(r)\dagger},
e^{-\omega_n\tau+in\theta_{(r)}}b_n^{(r)\dagger}\right)$ for all
creation operators which lie to the right of the projector. Then,
after the matrix element is computed, we multiply it by
$e^{E_0\tau}$ and integrate over $\tau$ and $\theta_r$.  Any
potential divergences come from the region near $\tau=0$. }.

Equation (\ref{okok}) can be written as \be =-{1\over 8} \left<
Q_{3\beta_1\dot\beta_2}\frac{P}{E_0-H_2}\left[H_2,Q_3^{\beta_1\dot\beta_2}
\right]\right> -{1\over 8} \left<
Q_{3\dot\beta_1\beta_2}\frac{P}{E_0-H_2}\left[H_2,Q_3^{\dot\beta_1\beta_2}
\right]\right> \label{com} \ee Up to order $g_2$ the following
equation holds \be
\left[H_2,Q_3^{\beta_1\dot\beta_2}\right]=\left[Q_2^{\beta_1\dot\beta_2},H_3\right]
\ee so that (\ref{com}) becomes \be ={1\over 8} \left<
Q_{3\beta_1\dot\beta_2}\frac{P}{E_0-H_2}\left[H_3,Q_2^{\beta_1\dot\beta_2}\right]
\right> +{1\over 8} \left<
Q_{3\dot\beta_1\beta_2}\frac{P}{E_0-H_2}\left[H_3,Q_2^{\dot\beta_1\beta_2}\right]
\right> \ee Since $Q_2$ commutes with $H_2$ one has \bea
 =&+&{1\over 8} \left<
Q_{2\beta_1\dot\beta_2}Q_3^{\beta_1\dot\beta_2}\frac{P}{E_0-H_2}
H_3\right> +{1\over 8} \left<
Q_{2\dot\beta_1\beta_2}Q_3^{\dot\beta_1\beta_2}\frac{P}{E_0-H_2}
H_3\right> \cr &+&{1\over 8} \left<
Q_{3\beta_1\dot\beta_2}\frac{P}{E_0-H_2} H_3
Q_2^{\beta_1\dot\beta_2}\right> +{1\over 8} \left<
Q_{3\dot\beta_1\beta_2}\frac{P}{E_0-H_2} H_3
Q_2^{\dot\beta_1\beta_2}\right> \cr &-&\left< H_3
\frac{P}{E_0-H_2} H_3\right> \eea and the last term cancels the
$H_3$ contribution to the energy shift. The final expression
for the energy shift is
\begin{eqnarray}\label{newnice}
\delta E^{(2)} =&+&{1\over 8} \left<
Q_{2\beta_1\dot\beta_2}Q_3^{\beta_1\dot\beta_2}\frac{P}{E_0-H_2}
H_3\right> +{1\over 8} \left<
Q_{2\dot\beta_1\beta_2}Q_3^{\dot\beta_1\beta_2}\frac{P}{E_0-H_2}
H_3\right> \cr &+&{1\over 8} \left<
Q_{3\beta_1\dot\beta_2}\frac{P}{E_0-H_2} H_3
Q_2^{\beta_1\dot\beta_2}\right> +{1\over 8} \left<
Q_{3\dot\beta_1\beta_2}\frac{P}{E_0-H_2} H_3
Q_2^{\dot\beta_1\beta_2}\right> \cr &+&{1\over 4} \left<
Q_{2\beta_1\dot\beta_2}Q_4^{\beta_1\dot\beta_2} \right> +{1\over
4} \left< Q_{2\dot\beta_1\beta_2}Q_4^{\dot\beta_1\beta_2} \right>
\cr &+&{1\over 4} \left< Q_{4\beta_1\dot\beta_2}
Q_2^{\beta_1\dot\beta_2}\right> +{1\over 4} \left<
Q_{4\dot\beta_1\beta_2} Q_2^{\dot\beta_1\beta_2}\right>
\end{eqnarray}

 It
is amusing to note that the vanishing energy correction for a
supersymmetric external state is manifest in (\ref{newnice}),
since if $Q_2$ annihilates the external state, all of the terms are
identically zero. Please note the discussion above equation (\ref{contact}),
where it is explained that for calculations $Q_4$ is set to zero.

Using the $\left|[{\bf 9}, {\bf 1}]\ra^{(ij)}\right.$ external state, we can check that what
is left is manifestly
convergent for the four impurity channel, and then show that the addition of impurities
will not disturb this, leaving $ {\cal O}(\lambda')$ contributions at every order in impurities.
We have two sorts of terms in (\ref{newnice}), which we can represent schematically
as follows,

\begin{equation}
\label{newstuf}
\delta E_1 = \sum_I \frac{ \Big( \la \Phi | \la I | Q_3 \ra \Big) \Big(  \la \Psi | \la I | H_3 \ra \Big)^* }
{E_{\Phi} - E_{I}} \qquad
\delta E_2 = \sum_I\frac{ \Big( \la \Phi | \la I | H_3 \ra \Big) \Big(  \la \Psi | \la I | Q_3 \ra \Big)^* }
{E_{\Phi} - E_{I}}
\end{equation}
where $|\Phi\ra$ is the $\left|[{\bf 9}, {\bf 1}]\ra^{(ij)}\right.$ external state,
$|\Psi\ra = Q_2 |\Phi\ra $, and $|I\ra$ is a level-matched, two-string intermediate state.
In order to evaluate the convergence and large $\mu$ behavior of these
terms, we can be entirely schematic. We take,
\begin{equation}
| \Psi \ra \sim \sqrt{-\mu\alpha_3} \,\b^\dag_n \, \a^\dag_{-n} | \wt\a_3 \ra
\qquad | \Phi \ra \sim \a^\dag_n \, \a^\dag_{-n} | \wt\a_3 \ra
\end{equation}
while for the purpose of evaluating convergence we can take
\begin{equation}\label{conv}
G^{(1)}_p \sim \frac{1}{\sqrt{p}} \qquad K^{(1)}_{-p} \sim \mbox{constant} \qquad
{\widetilde N}^{3\,r}_{n\, p} \sim \frac{1}{p} \qquad
{\widetilde N}^{r\,s}_{q\, p} \sim \frac{1}{p + q}
\end{equation}
where we take all integers to be positive. Let us begin with
the first type of term in (\ref{newstuf}), we have two choices for
four impurity intermediate states,
\bea
\label{inter}
&&|I\ra \sim \a^\dag_{p_1}\b^\dag_{p_2}\a^\dag_{p_3}\a^\dag_{p_4} | \wt\a_1 \ra | \wt\a_2 \ra \cr
&&|I\ra \sim \a^\dag_{p_1}\b^\dag_{p_2}\b^\dag_{p_3}\b^\dag_{p_4} | \wt\a_1 \ra | \wt\a_2 \ra
\eea

We can proceed with the first one, which will give,
\bea
&&\delta E_1 \sim
\sqrt{x}
\sum_{p_1\,p_2\,p_3\,p_4}\frac{1}
{2\,r\,\omega_n - \sum_{i=1}^4\,\omega_{p_i}}\,\cdot\cr
&&
\la \wt\a_3 | \a_n \, \a_{-n} \la \wt\a_2 | \la \wt\a_1
| \a_{p_1}\, \b_{p_2} \,\a_{p_3}\, \a_{p_4}\, |Q_3\ra
\Big( \la \wt\a_3 | \b_n \, \a_{-n} \la \wt\a_2 | \la \wt\a_1
| \a_{p_1}\, \b_{p_2} \,\a_{p_3}\, \a_{p_4}\, |H_3\ra \Big)^*~~~~~~
\eea
where $x = -\mu\alpha_3$ and $\sum_i p_i=0$. Before continuing with
contractions we should note that because of the appearance of
multiple fermionic oscillators we should be using the forms of
$|H_3\ra$ and $|Q_3\ra$ given in Appendix \ref{dynamical}. We refer the reader there
for these expressions. There are two general ways in which we can
contract the $\b^{(r)}$'s. They can connect to factors of $\sum_m
G_m \b^\dag_m$ in the prefactors of $|H_3\ra$ and $|Q_3\ra$, or they can
pair-up to bring down a factor of ${\widetilde Q}^{r\,s}_{m\, p} -
{\widetilde Q}^{r\,s}_{p\, m}$ from $|E_{\beta}\rangle$ (see Appendix \ref{appA}).
As far as convergence and large $x$ power-counting is concerned however,
$G^{(r)}_m G^{(s)}_p$ is equivalent to ${\widetilde Q}^{r\,s}_{m\,
p} - {\widetilde Q}^{s\,r}_{p\, m}$, and so we will simply use the
former. When contracting $\b^{(3)}$'s there is a fundamental
difference between $G^{(3)}_n G^{(r)}_p$ and ${\widetilde
Q}^{3\,r}_{n\, p} - {\widetilde Q}^{r\,3}_{p\, n}$, as far as large
$x$ behavior is concerned, because of the pole in the latter. In
fact ${\widetilde Q}^{3\,r}_{n\, p} - {\widetilde Q}^{r\,3}_{p\, n}$
is essentially equivalent to ${\widetilde N}^{3\,r}_{n\, p}$ and
therefore the two can be interchanged in this analysis.

Because $K_{-p}$ goes as a constant for large $p$, the worst
convergence will always be realized by contracting the
intermediate bosonic impurities with the prefactors of $|H_3\ra$ and
$|Q_3\ra$. These contractions will yield\footnote{ Note that any
contraction which would yield a delta function on the external
state's spacetime indices is naturally zero here because we have
chosen to analyze the traceless symmetric $|[{\bf 9}, {\bf
1}]\ra^{(ij)}$ state. It is a simple matter to analyze the trace
state of section \ref{trace} here, and one finds convergence as
well, however the number of (inverse) powers of summation
variables will be 4 in the worst case, and thus the convergence is
marginal (see discussion below (\ref{meme})). In no case does 
$\sqrt{\lambda'}$ behavior occur here.},

\bea
\label{bb}
&&\delta E_1 \sim
\sqrt{x}\sum_{p_1\,p_2\,p_3\,p_4 }
\frac{
G^{(1)}_{p_2} {\wt N}^{3\,1}_{-n\, p_1} K^{(1)}_{-p_3} {\wt N}^{3\,1}_{n\, p_4}
\times  K^{(1)}_{-p_3} K^{(1)}_{p_4} {\wt N}^{3\,1}_{-n\, p_1}
\cases{
{\widetilde Q}^{3\,1}_{n\, p_2} - {\widetilde Q}^{1\,3}_{p_2\, n} \cr
G^{(3)}_n G^{(1)}_{p_2}
}}
{2\,r\,\omega_n - \sum_{i=1}^4\,\omega_{p_i}}
\eea
Taking $p_4 = -(p_1+p_2+p_3)$, and using (\ref{conv}) we see that,

\bea
&&\delta E_1 \sim \sum_{p_1\,p_2\,p_3} \frac{1}{(p_1+p_2+p_3)^2} \frac{1}{p_1^2}
\cases{
\frac{1}{p_2^{3/2}}\cr
\frac{1}{p_2}}
\eea
where all $p_i$ are considered absolute valued, or equivalently the
sum considered over positive integers. This is manifestly convergent. Continuing
on to evaluate the leading $x$ dependence, for the top choice in (\ref{bb})
we have poles for all three summation
variables, while in the large $x$ limit the $K$'s go as constants,
$G\sim 1/\sqrt{x}$ and the energy denominator is linear in $x$, thus giving
$\delta E_1 \sim 1/x$. For the bottom choice in (\ref{bb}), $p_1$ and $p_3$
have poles, while the sum over $p_2$ must be executed using (\ref{nopoles}).
The scaling turns out identical however. Thus $\delta E_1/\mu$ is convergent
and ${\cal O}(\lambda')$. One can repeat this argumentation for the second
intermediate state in (\ref{inter}) and find the same behavior. Also the
entire exercise may be repeated for $\delta E_2$ using the following
intermediate states,

\bea\label{meme}
&&|I\ra \sim \a^\dag_{p_1}\a^\dag_{p_2}\a^\dag_{p_3}\a^\dag_{p_4} | \wt\a_1 \ra | \wt\a_2 \ra \cr
&&|I\ra \sim \a^\dag_{p_1}\a^\dag_{p_2}\b^\dag_{p_3}\b^\dag_{p_4} | \wt\a_1 \ra | \wt\a_2 \ra
\eea
and one discovers the same behavior. The essential point is that we will
always have at least 5 (inverse) powers of the summation variables, while the number of summation
variables is 3. Alternate positionings of the oscillators in the intermediate
states such as
$ |I\ra \sim \a^\dag_{p_1}\a^\dag_{p_2} | \wt\a_1 \ra \, \a^\dag_{p_3}\a^\dag_{p_4}  | \wt\a_2 \ra $
only improves the convergence, since level matching removes one more summation variable
in these cases.

We can now consider adding additional pairs of fermionic and
bosonic impurities to the intermediate state $|I\ra$. This will
add two factors of ${\wt N}^{1\,1}_{p_i\,p_j}$ or two factors of
$G^{(1)}_{p_i} G^{(1)}_{p_j}$ (or equivalently two factors of $\wt Q^{1\,1}_{p_i p_j}
- \wt Q^{1\,1}_{p_j p_i}$). Either way the number of powers of summation
variables increases in concert with the number of summation
variables, preserving the convergence. Similarly the leading
behavior in $\lambda'$ is unaffected. So it would seem that there
are ${\cal O}(\lambda')$ contributions to $\delta E^{(2)}/\mu$ at
every order in impurities, however any non-perturbative
$\sqrt{\lambda'}$ behavior is absent.

\section{Conclusions}

In this paper we have discovered logarithmic divergences in the
one-loop mass shift of two-impurity string states on the plane
wave background. As superstring amplitudes should be finite,
these divergences ought to cancel, and we find that they do,
via a cancellation between the $H_3$ vertex and the contact term.
This is reminiscent of similar results for string amplitudes
in Minkowski space.

Further we have shown that the apparent non-perturbative
$\sqrt{\lambda'}$ behavior of contributions to the mass shift
from an impurity non-conserving channel (where the number of
impurities is increased by two in the intermediate states)
is in fact an artifact of these logarithmic divergences and 
vanishes with them, leaving an $\cal O(\lambda')$ contribution.
We have also given arguments which generalize the above statements
to intermediate states with an arbitrary number of impurities,
and up to a possible role played by an as yet unknown $Q_4$, have
derived a formula (equation (\ref{newnice})) for the mass shift which appears to be
manifestly devoid of any $\sqrt{\lambda'}$ or non-convergent 
behavior.   

We have also shown that generically, every order in impurities
contributes an $\cal O(\lambda')$ piece to the mass shift, making
the prospects for computing this quantity, so that it can be
matched to Yang-Mills theory, rather disappointing. On the other
hand it is heartening that the string amplitudes appear to match
the Yang-Mills results in terms of the leading power of $\lambda'$,
for any intermediate state.

A formal evaluation of the expectation values in equation
(\ref{newnice}) may be possible, and we hope to publish
further work in this direction soon.

\section*{Acknowledgments}

One of the authors, G.W.S., thanks the Yukawa Institute for Theoretical
Physics at Kyoto University. Discussions during the YITP workshop
YITP-W-05-08 on ``String Theory and Quantum Field Theory" were useful for
completion of this work.
Another author, M.O., thanks Paolo Di Vecchia for useful discussions.
This work was partially supported by NSERC of Canada, the String Theory
Collaborative Research Group of the Pacific Institute for Mathematical
Sciences and the Strings and Particles Collaborative Research Team of the
Pacific Institute for Theoretical Physics.

\appendix{\section{Free string on the pp-wave}\label{freestrings}}

The light-cone action in the pp-wave background is
\begin{eqnarray}
S_b = \frac{e(\alpha)}{4\pi\alpha'}\int d\tau
\int_0^{2\pi\vert\alpha\vert} d\sigma\left( {\partial_\tau
X}^I{\partial_\tau X}^I-{\partial_\sigma X^I}{\partial_\sigma
X^I}-\mu^2X^IX^I\right)+\nonumber \\ +\frac{1}{8\pi}\int
d\tau\int_0^{2\pi|\alpha|}d\sigma\left(i\bar\vartheta\partial_\tau\theta+
i\vartheta\partial_\tau\bar\vartheta-\vartheta\partial_\sigma\bar\vartheta
+\bar\vartheta\partial_\sigma\vartheta-2\mu\bar\vartheta\Pi\vartheta\right)\end{eqnarray}
where $I=1,...,8$, $e(\alpha)={\rm sign}(\alpha)$, $\alpha =
\alpha'p^+$, $\theta$ is an 8-component positive chirality spinor of
SO(8) and $\Pi=\Gamma^1\Gamma^2\Gamma^3\Gamma^4$ is a symmetric,
traceless projection operator, $\Pi^2=1$.  We use the convention
that $p^+<0$ for incoming strings and $p^+>0$ for outgoing strings.
The mode expansions for the bosonic and ferminonic coordinates
and their conjugate momenta are
\begin{eqnarray}
&&X^I(\sigma) = x_0^I+\sqrt{2}\sum_{n=1}^\infty \left(
x_n^I\cos\frac{n\sigma}{|\alpha|} + x_{-n}^I
\sin\frac{n\sigma}{|\alpha|}\right) \\
&&P^I(\sigma) =
\frac{1}{2\pi|\alpha|}\left[p_0^I+\sqrt{2}\sum_{n=1}^\infty \left(
p_n^I\cos\frac{n\sigma}{|\alpha|} + p_{-n}^I
\sin\frac{n\sigma}{|\alpha|}\right)\right]\\
&&\vartheta^a(\sigma) = \vartheta_0^a+\sqrt{2}\sum_{n=1}^\infty \left(
\vartheta_n^a\cos\frac{n\sigma}{|\alpha|} + \vartheta_{-n}^a
\sin\frac{n\sigma}{|\alpha|}\right) \\
&&\lambda^a(\sigma) =\frac
{1}{2\pi|\alpha|}\left[ \lambda^a_0+\sqrt{2}\sum_{n=1}^\infty \left(
\lambda_n^a\cos\frac{n\sigma}{|\alpha|} + \lambda_{-n}^a
\sin\frac{n\sigma}{|\alpha|}\right)\right]
\end{eqnarray}
where $2 \lambda_n^a = |\alpha|\bar\vartheta_n^a$ and $a$ is an $SO(8)$ spinor index.
The non-vanishing (anti-)commutators of the Fourier modes are
\begin{equation}
\left[x_m^I,p_n^J\right]=i\delta^{IJ}\delta_{mn} ~~,~~
\left\{\vartheta^a_m,
\lambda^b_n\right\}=\delta^{ab}\delta_{mn}\end{equation}  and lead
to \begin{equation} \left[
x^I(\sigma),p^J(\sigma')\right]=i\delta^{IJ}\delta(\sigma-\sigma')
~~,~~ \left\{
\vartheta^a(\sigma),\lambda^b(\sigma')\right\}=\delta^{ab}\delta(\sigma
- \sigma')\end{equation}
The modes can also be written in terms of oscillators as
\begin{equation} x_n^I = i\sqrt{\frac{\alpha}{2\omega_n}}\left(
a^I_n - a^{I\dagger}_n\right) ~~,~~ p_n^I =
i\sqrt{\frac{\alpha}{2\omega_n}}\left(a^I_n +
a^{I\dagger}_n\right) ~~,~~ \left[a^I_m,
a^{J\dagger}_n \right] = i\delta^{IJ}\delta_{mn}
\end{equation}
\begin{eqnarray}
&&\vartheta^a_n = \frac{c_n}{\sqrt{\vert\alpha\vert}}\left[
\left(1+\rho_n\Pi\right)b^a_n+e(n\alpha)\left(
1-\rho_n\Pi\right)b_{-n}^{a\dagger}\right]\label{fnm}\\
&&\lambda^a_n =
\frac{\sqrt{\vert\alpha\vert}c_n}{2}\left[\left(
1+\rho_n\Pi\right)b_{n}^{a\dagger}+e(n\alpha)
\left(1-\rho_n\Pi\right)b^a_{-n}\right]\label{fnmm}\\
&&\left\{ b_m^a,b_n^{b\dagger}\right\}=\delta^{ab}\delta_{mn}
\end{eqnarray}
with $\omega_n = \sqrt{ n^2 + (\mu\alpha)^2}$,
$\rho_n=\frac{\omega_n - |n|}{\mu\alpha}$,
$c_n=\frac{1}{\sqrt{1+\rho_n^2}}$.

The free string Hamiltonian for the $r$-th string
\begin{eqnarray}
 &&\hspace{-1cm}H_2^{(r)}=
\frac{1}{2}\int_0^{2\pi|\a_r|}d\sigma\left[
2\pi\alpha' P^{(r)2}+\frac{1}{2\pi\alpha'}(\partial_\sigma X^{(r)})^2
+\frac{1}{2\pi\alpha'}\mu^2X^{(r)2})\right]\\
&&+\frac{1}{2}\int_{0}^{2\pi|\a_r|}d\sigma\left[
-2\pi \a' \lambda^{(r)}\partial_\sigma \lambda^{(r)}+
\frac{1}{2\pi\a'}\theta^{(r)}\partial_\sigma\theta^{(r)}
+2\mu\lambda^{(r)}\Pi\theta^{(r)}\right]\nonumber
\end{eqnarray}
in this Fock space basis reduces to
\begin{eqnarray}
 H^{(r)}_2=\sum_{n=-\infty}^{\infty}
\frac{\omega_n^{(r)}}{|\a_r|}
\left(
a_n^{(r)\dagger}a_n^{(r)}+b_n^{(r)\dagger}b_n^{(r)}
\right).
\label{h2ab}
\end{eqnarray}

Isometries of the pp-wave background are generated by $H$, $P^+$, $J^{+I}$,
$J^{ij}$ and $J^{i'j'}$ where $i,j=1,2,3,4$, $i'j'=5,6,7,8$.
The latter two are angular momentum generators of the transverse $SO(4)\times SO(4)$ symmetry.
There are 32 conserved supercharges $Q^+$, $\bar{Q}^+$ and $Q^-$, $\bar{Q}^-$.
These generators are divided into two groups, kinematical generators
$$
P^I~,~P^+~,~J^{+I}~,~J^{ij}~,~J^{i'j'}~,~Q^+~,~\bar Q^+
$$
which are not corrected when string interactions are
introduced and the dynamical generators
$$
H~,~ Q^-~,~\bar Q^-
$$
which get corrections from interactions.
The quadratic parts of $H$
is given in (\ref{h2ab}) above and the supercharges are given by
\bea
\label{Q+}
&&Q^+_{(r)} = \sqrt{\frac{2}{\a'}}\int_0^{2\pi|\a_r|}d\s_r\,\sqrt{2}\l_r\,,\\
\label{qfield}
&&Q^-_{(r)} =\sqrt{\frac{2}{\a'}}\int_0^{2\pi|\a_r|}d\s_r\,\left[2\pi\a'e(\a_r)p_r\g\l_r-ix'_r\g\bar{\l}_r-i\m x_r\g\Pi\l_r\right]\,,
\eea
$\bar{Q}^{\pm}_{(r)}=e(\a_r)\bigl[Q_{(r)}^{\pm}\bigr]^{\dag}$
and $\gamma^I$ are the $SO(8)$ Weyl matrices.\footnote{ The
$SO(8)$ gamma-matrices are $\Gamma^I=\left(\matrix{ 0 &
\gamma^I \cr \bar \gamma^I & 0 \cr}\right)$.}

The mode expansion of $Q^-$ is
\bea\label{q-mode}
Q^-_{(r)}& =&\frac{e(\a_r)}{\sqrt{|\a_r|}}\g
\Bigl(\sqrt{\m}\left[a_{0(r)}(1+e(\a_r)\Pi)+a_{0(r)}^{\dag}(1-e(\a_r)\Pi)\right]\l_{0(r)}\cr
&+&\sum_{n\neq 0}\sqrt{|n|}\left[a_{n(r)}P_{n(r)}^{-1}b_{n(r)}^{\dag}
+e(\a_r)e(n)a_{n(r)}^{\dag}P_{n(r)}b_{-n(r)}\right]\Bigr)\,,
\eea
where
\begin{equation}
P_{n(r)}\equiv\frac{1-\r_{n(r)}\Pi}{\sqrt{1-\r_{n(r)}^2}}
=\frac{1+\Pi}{2}U_{|n|(r)}^{1/2}+\frac{1-\Pi}{2}U_{|n|(r)}^{-1/2}\,,\qquad
U_{n(r)}\equiv\frac{\o_{n(r)}-\m\a_r}{n}\,.
\end{equation}

These operators generate the superalgebra
\bea\label{comm}
&&[H,P^I] = -i\m^2J^{+I}\,,\qquad [H,Q^+]=-\m\Pi Q^+\,,\\
&&\{Q^-_{\da},\bar{Q}^-_{\db}\} =
2\d_{\dot{a}\dot{b}}H-i\m\bigl(\g_{ij}\Pi\bigr)_{\dot{a}\dot{b}}J^{ij}+i\m\bigl(\g_{i'j'}\Pi\bigr)_{\dot{a}\dot{b}}J^{i'j'}\,.
\eea

The fermionic normal modes (\ref{fnm},\ref{fnmm}) break the $SO(8)$ symmetry to $SO(4)\times SO(4)$.
To make this symmetry manifest it is convenient to label representations of $SO(4)_1\times SO(4)_2$
through $(SU(2)\times SU(2))_1\times (SU(2)\times SU(2))_2$ spinor indices.
With this decomposition of the R-charge index,
the fermionic fields
$\vartheta^a$ and $\lambda^a$,
are expressed in terms of creation operators
$b^{\dagger}_{\a_1\a_2}$ and $b^{\dagger}_{\dot\a_1\dot\a_2}$ which transform
in the  $(1/2,0,1/2,0)$
and $(0,1/2,0,1/2)$ representations of  $(SU(2)\times SU(2))_1\times (SU(2)\times SU(2))_2$, respectively;
$\alpha_k$,$\da_k$ being two-component Weyl indices of $SO(4)_k$.

The $SO(8)$ vector index $I$ splits into two $SO(4)\times SO(4)$ vector indices
$(i,i')$ so that we use vector index $i=1,\dots,4$ and bi-spinor indices  $\a_1,\da_1=1,2$
for the first $SO(4)$ and $(i',\a_2,\da_2)$ for the second $SO(4)$.
Vectors are constructed in terms of bi-spinor indices as
$(a_n)_{\a_1\dot \a_1}=\sigma^i_{\a_1\dot\a_1} a^i_n/\sqrt{2}$,
$(a_n)_{\a_2\dot \a_2}=\sigma_{\a_2\da_2}^{i'} a^{i'}_n/\sqrt{2}$ and transform as
$(1/2,1/2,0,0)$ and $(0,0,1/2,1/2)$, respectively. Here the $\s$-matrices
consist of the usual Pauli-matrices together with the 2d unit matrix
\be
\s^i_{\a\da}=\bigl(i\t^1,i\t^2,i\t^3,-1\bigr)_{\a\da}
\ee
and satisfy the
reality properties $\bigl[\s^i_{\a\da}\bigr]^{\dag} = {\s^i}^{\da\a}$\,,
$\bigl[{\s^i}_{\a}^{\da}\bigr]^{\dag} = -{\s^i}^{\a}_{\da}$.

Spinor indices are raised and lowered with the two-dimensional Levi-Civita symbols,
$\e_{\a\b}=\e_{\da\db}\equiv\left(\matrix{ 0
& 1 \cr -1 & 0 }\right)$, for example
\begin{equation}
\s^i_{\a\da} = \e_{\a\b}\e_{\da\db}{\s^i}^{\db\b}
\equiv \e_{\a\b}{\s^i}^{\b}_{\da} \equiv \e_{\da\db}{\s^i}^{\db}_{\a}\,.
\end{equation}

The $\s$-matrices satisfy the relations
\begin{equation}
\s^i_{\a\da}{\s^j}^{\da\b}+\s^j_{\a\da}{\s^i}^{\da\b}
=2\d^{ij}\d_{\a}^{\b}\,,\qquad
{\s^i}^{\da\a}\s^j_{\a\db}+{\s^j}^{\da\a}\s^i_{\a\db}=
2\d^{ij}\d^{\da}_{\db}\,.
\end{equation}
Some other properties satisfied by these matrices are
\bea
\label{rel1}
&&\e_{\a\b}\e^{\g\d}  = \d_{\a}^{\d}\d_{\b}^{\g}-\d_{\a}^{\g}\d_{\b}^{\d}\,,\\
&&\s^i_{\a\db}{\s^j}^{\db}_{\b}  = -\d^{ij}\e_{\a\b}+\s^{ij}_{\a\b}\,,\qquad
(\s^{ij}_{\a\b}\equiv \s^{[i}_{\a\da}{\s^{j]}}^{\da}_{\b}=\s^{ij}_{\b\a})\\
&&\s^i_{\a\da}{\s^j}^{\a}_{\db} = -\d^{ij}\e_{\da\db}+\s^{ij}_{\da\db}\,,
\qquad (\s^{ij}_{\da\db}\equiv \s^{[i}_{\a\da}{\s^{j]}}^{\a}_{\db}=\s^{ij}_{\db\da})\\
&&\s^k_{\a\da}\s^{k}_{\b\db}  = 2\e_{\a\b}\e_{\da\db}\,,\\
&&\s^{kl}_{\a\b}\s^{kl}_{\g\d}  = 4(\e_{\a\g}\e_{\b\d}+\e_{\a\d}\e_{\b\g})\,,\\
&&\s^{kl}_{\a\b}\s^{kl}_{\dg\dd}  = 0\,,\\
\label{rel7}
&&2\s^i_{\a\da}\s^{j}_{\b\db} = \d^{ij}\e_{\a\b}\e_{\da\db}
+\s^{k(i}_{\a_1\b_1}\s^{j)k}_{\da_1\db_1}
-\e_{\a\b}\s^{ij}_{\da\db}-\s^{ij}_{\a\b}\e_{\da\db}\,.
\eea

In this basis the gamma matrices have the following representation
\bea
&&\gamma^i_{a\dot{a}} =
\left(\matrix{
0 & \s^i_{\a_1\db_1}\d_{\a_2}^{\b_2} \cr {\s^i}^{\da_1\b_1}\d^{\da_2}_{\db_2} & 0}
\right)\ ,\qquad~~
\g^i_{\dot{a}a} =\left(
\matrix{
0 & \s^i_{\a_1\db_1}\d^{\da_2}_{\db_2} \cr {\s^i}^{\da_1\b_1}\d_{\a_2}^{\b_2} & 0}
\right)\ ,\\
&&\g^{i'}_{a\dot{a}} =\left(
\matrix{
-\d_{\a_1}^{\b_1}\s^{i'}_{\a_2\db_2} & 0 \cr0 & \d^{\da_1}_{\db_1}{\s^{i'}}^{\da_2\b_2}}
\right)\ ,\qquad~~
\g^{i'}_{\dot{a}a} =\left(
\matrix{
-\d_{\a_1}^{\b_1}{\s^{i'}}^{\da_2\b_2} & 0 \cr 0 & \d^{\da_1}_{\db_1}\s^{i'}_{\a_2\db_2}}
\right)\ .
\eea
and the projector reads
\begin{equation}
\Pi_{ab} =
\left(\matrix{\bigl(\s^1\s^2\s^3\s^4\bigr)_{\a_1}^{\b_1}\d_{\a_2}^{\b_2} & 0 \cr
0 & \bigl(\s^1\s^2\s^3\s^4\bigr)^{\da_1}_{\db_1}\d^{\da_2}_{\db_2}}\right)
=
\left(\matrix{
\d_{\a_1}^{\b_1}\d_{\a_2}^{\b_2} & 0 \cr 0 & -\d^{\da_1}_{\db_1}\d^{\da_2}_{\db_2}}\right)\,,
\end{equation}
so that $(1\pm\Pi)/2$ projects onto
$(1/2,0,1/2,0)$ and $(0,1/2,0,1/2)$, respectively.

  The supercharge  $Q^-_{\a_1\dot
\b_2}$ is a $(1/2,0,0,1/2)$ and $Q^-_{\dot\a_1\b_2}$ is a $(0,1/2,1/2,0)$
representation. In this notation it is convenient to
 define the linear combinations of the free supercharges
\be
\sqrt{2}\eta Q\equiv Q^-+i\bar{Q}^-~~~,~~~\sqrt{2}\bar{\eta} \widetilde{Q}
\equiv Q^--i\bar{Q}^-
\ee
where $\eta=e^{i\pi/4}$.
On the space of physical state they satisfy the dynamical constraints
\bea
&&\left\{Q_{\a_1\da_2},Q_{\b_1\db_2}\right\}=
\left\{\widetilde{Q}_{\a_1\da_2},\widetilde{Q}_{\b_1\db_2}\right\}=
-2\epsilon_{\a_1\b_1}\epsilon_{\da_2\db_2}H\cr
&&\left\{Q_{\a_1\da_2},\widetilde{Q}_{\b_1\db_2}\right\}=
-\mu \epsilon_{\da_2\db_2}\left(\sigma^{ij}\right)_{\a_1\b_1}J^{ij}+
\mu \epsilon_{\a_1\b_1}\left(\sigma^{i'j'}\right)_{\da_2\db_2}J^{i'j'}
\label{dynamconst}
\eea
and similarly for $Q_{\da_1\a_2}$ and $\widetilde{Q}_{\db_1\b_2}$.

Among states that are created by two oscillators, the state with
quantum numbers $(1,1,0,0)$ and $(0,0,1,1)$ which are created by two
bosons have no analogs amongst the two oscillator states containing
either one or two fermions.  Thus, they are not mixed with other
members of the supermultiplet.  These states in the main text are denoted
$\left|[{\bf 9}, {\bf 1}]\ra^{(ij)}\right.$ and
$|[{\bf 1}, {\bf 9}]\ra^{(i'j')}$
in SO(8) notation.

\section{Solving the dynamical constraints}
\label{dynamical}

When string interactions are considered, the dynamical generators of the superalgebra
receive $g_2$ corrections
so that they can be generally written in a perturbative $g_2$ expansion
\bea
&&H=H_2+g_2H_3+g^2_2H_4+\dots\ ,\cr
&&Q_{\a_1\da_2}=Q_{2\a_1\da_2}+g_2Q_{3\a_1\da_2}+g^2_2 Q_{4 \a_1\da_2} + \dots
\label{H}
\eea
$H_3,Q_3$ are the operators representing a three string interaction and
$H_4,Q_4$ are contact term interactions.
As we shall see $H_4$ is induced by cubic supercharges.
Such an expansion can be used to solve perturbatively
the dynamical constraints (\ref{dynamconst})
At order ${\mc O}(g_2)$ the dynamical constraints (\ref{dynamconst}) become
\bea
&&\{Q_{2\,\a_1\da_2},Q_{3\,\b_1\db_2}\}+\{Q_{3\,\a_1\da_2},Q_{2\,\b_1\db_2}\} = -2\e_{\a_1\b_1}\e_{\da_2\db_2}H_3\,,\label{dyn1}\\
&&\{\wt{Q}_{2\,\a_1\da_2},\wt{Q}_{3\,\b_1\db_2}\}+\{\wt{Q}_{3\,\a_1\da_2},\wt{Q}_{2\,\b_1\db_2}\}
= -2\e_{\a_1\b_1}\e_{\da_2\db_2}H_3\,,\label{dyn2}\\
&&\{Q_{2\,\a_1\da_2},\wt{Q}_{3\,\b_1\db_2}\} + \{Q_{3\,\a_1\da_2},\wt{Q}_{2\,\b_1\db_2}\} = 0\label{dyn3}\,.
\eea
It is convenient~\cite{Cremmer:1974jq} to express $H_3$ and $Q_3$ as states $|H_3\rangle$ and $|Q_3\rangle$ in the
multi-string Hilbert space and work in the number basis where the dynamical generators can
be written as ${\cal P}|V\rangle$, where ${\cal P}$ are prefactors determined by imposing the dynamical
constraints and $|V\rangle$ is the kinematical part of the vertex and implements the continuity conditions.
Equations (\ref{dyn1}-\ref{dyn3}) become
\bea
\label{1}
&&\sum_{r=1}^3 Q_{(r)\,\a_1\da_2}|Q_{3\,\b_1\db_2}\ra+\sum_{r=1}^3 Q_{(r)\,\b_1\db_2}|Q_{3\,\a_1\da_2}\ra
= -2\e_{\a_1\b_1}\e_{\da_2\db_2}|H_3\ra\,,\\
\label{2}
&&\sum_{r=1}^3Q_{(r)\,\da_1\a_2}|Q_{3\,\db_1\b_2}\ra+\sum_{r=1}^3 Q_{(r)\,\db_1\b_2}|Q_{3\,\da_1\a_2}\ra  = 
-2\e_{\da_1\db_1}\e_{\a_2\b_2}|H_3\ra\,,\\
\label{3}
&&\sum_{r=1}^3Q_{(r)\,\a_1\da_2}|Q_{3\,\db_1\b_2}\ra+\sum_{r=1}^3 Q_{(r)\,\db_1\b_2}|Q_{3\,\a_1\da_2}\ra = 0\,,
\eea
and analogously for $Q\to \widetilde{Q}$.

Making an ansatz for, say $Q_{3\,\a_1\da_2}$,
compatible with  the requirement that the Hamiltonian prefactor in
its functional form is quadratic in derivatives~\cite{Pankiewicz:2003kj},
into~(\ref{1}) and demanding that the result only involves the tensor $\e_{\a_1\b_1}\e_{\da_2\db_2}$
fixes $Q_{3\,\a_1\da_2}$ and consequently also $H_3$ up to their normalization.
The same procedure applies to $\wt{Q}_{3\,\a_1\da_2}$ and requires that its normalization is the
same as of $Q_{3\,\a_1\da_2}$.\\
A possible choice for the three-string vertex and dynamical supercharges~\footnote{ 
The three-string vertex and dynamical supercharges in the open string case have been constructed in~\cite{Chandrasekhar:2003fq}.}
that solves the dynamical constraints up to order $g_2$ is~\cite{Pankiewicz:2003kj}
\bea
|H_3\ra =&&
-g_2\,f(\m\a_3\,,\,\frac{\a_1}{\a_3})\frac{\alpha'}{8\,\a_3^3}
\Bigl[\bigl(K_i\K_j-\frac{\m\k}{\a'}\d_{ij}\bigr)v^{ij}
-\bigl(K_{i'}\K_{j'}-\frac{\m\k}{\a'}\d_{i'j'}\bigr)v^{i'j'}
\nonumber\\
&&-K^{\da_1\a_1}\K^{\da_2\a_2}s_{\a_1\a_2}(Y)s^*_{\da_1\da_2}(Z)
-\K^{\da_1\a_1}K^{\da_2\a_2}s^*_{\a_1\a_2}(Y)s_{\da_1\da_2}(Z)\Bigr]|V\ra\,,\cr
|Q_{3\,\b_1\db_2}\ra =&&
 g_2\,\eta\,f(\m\a_3\,,\,
\frac{\a_1}{\a_3})\frac{1}{4\, \a_3^3}\,\sqrt{-\frac{\a'\k}{2}}
\Bigl(s_{\dg_1\db_2}(Z)t_{\b_1\g_1}(Y)\K^{\dg_1\g_1}\cr
&&+ is_{\b_1\g_2}(Y)t^*_{\db_2\dg_2}(Z)\K^{\dg_2\g_2}\Bigr)|V\ra\,,\cr
|Q_{3\,\db_1\b_2}\ra =&& g_2\,\eta\,f(\m\a_3\,,\,
\frac{\a_1}{\a_3})\frac{1}{4\, \a_3^3}\,\sqrt{-\frac{\a'\k}{2}}
\Bigl(s^*_{\g_1\b_2}(Y)t^*_{\db_1\dg_1}(Z)\K^{\dg_1\g_1}
\cr
&&+is^*_{\db_1\dg_2}(Z)t_{\b_2\g_2}(Y)\K^{\dg_2\g_2}\Bigr)|V\ra\,.
\eea
where $\k\equiv\a_1\a_2\a_3$, $\a_3<0$
for the incoming and $\a_{1,2}>0$ for the outgoing strings,
$K^I,\ \wt{K}^I$ are defined in~(\ref{k}), $Y$ and $Z$ in
(\ref{yz}) and
\begin{equation}
\K^{\dg_1\g_1} \equiv \K^i{\s^i}^{\dg_1\g_1}\,,\qquad
\K^{\dg_2\g_2} \equiv \K^{i'}{\s^{i'}}^{\dg_2\g_2}\,,
\end{equation} Moreover
\bea
&&v^{ij}  =
\d^{ij}\Bigl[1+\frac{1}{12}\bigl(Y^4+Z^4\bigr)+\frac{1}{144}Y^4Z^4\Bigr]\nonumber\\
&&-\frac{i}{2}\Bigl[{Y^2}^{ij}\bigl(1+\frac{1}{12}Z^4\bigr)
-{Z^2}^{ij}\bigl(1+\frac{1}{12}Y^4\bigr)\Bigr]
+\frac{1}{4}\bigl[Y^2Z^2\bigr]^{ij}\,,\\
&&v^{i'j'}  =
\d^{i'j'}\Bigl[1-\frac{1}{12}\bigl(Y^4+Z^4\bigr)+\frac{1}{144}Y^4Z^4\Bigr]\nonumber\\
&&-\frac{i}{2}\Bigl[{Y^2}^{i'j'}\bigl(1-\frac{1}{12}Z^4\bigr)
-{Z^2}^{i'j'}\bigl(1-\frac{1}{12}Y^4\bigr)\Bigr]
+\frac{1}{4}\bigl[Y^2Z^2\bigr]^{i'j'}\,.
\eea
Here we defined
\begin{equation}
{Y^2}^{ij} \equiv \s^{ij}_{\a_1\b_1}{Y^2}^{\a_1\b_1}\,,\quad
{Z^2}^{ij} \equiv \s^{ij}_{\da_1\db_1}{Z^2}^{\da_1\db_1}\,,\quad
\bigl(Y^2Z^2\bigr)^{ij} \equiv {Y^2}^{k(i}{Z^2}^{j)k}
\end{equation}
and analogously for the primed indices.
We have also introduced the following 
quantities quadratic and cubic in $Y$ and symmetric in spinor indices
\begin{equation}\label{y2}
Y^2_{\a_1\b_1} \equiv Y_{\a_1\a_2}Y^{\a_2}_{\b_1}\,,\qquad Y^2_{\a_2\b_2} \equiv Y_{\a_1\a_2}Y^{\a_1}_{\b_2}\,,
\end{equation}
\begin{equation}\label{y3}
Y^3_{\a_1\b_2} \equiv Y^2_{\a_1\b_1}Y^{\b_1}_{\b_2}=-Y^2_{\b_2\a_2}Y^{\a_2}_{\a_1}\,,
\end{equation}
and quartic in $Y$ and antisymmetric in spinor indices
\begin{equation}
Y^4_{\a_1\b_1} \equiv Y^2_{\a_1\g_1}{Y^2}^{\g_1}_{\b_1}=-\frac{1}{2}\e_{\a_1\b_1}Y^4\,,\qquad
Y^4_{\a_2\b_2} \equiv Y^2_{\a_2\g_2}{Y^2}^{\g_2}_{\b_2}=\frac{1}{2}\e_{\a_2\b_2}Y^4\,,
\end{equation}
where
\begin{equation}\label{y4}
Y^4 \equiv Y^2_{\a_1\b_1}{Y^2}^{\a_1\b_1}=-Y^2_{\a_2\b_2}{Y^2}^{\a_2\b_2}\,.
\end{equation}
The spinorial quantities $s$ and $t$ are defined as
\begin{equation}
s(Y) \equiv Y+\frac{i}{3}Y^3\,\ ,~~~~~t(Y) \equiv \e+iY^2-\frac{1}{6}Y^4\,.
\end{equation}
Analogous definitions can be given for $Z$.
The other dynamical supercharges can be obtained from $|\wt{Q}\ra=|Q^*\ra$.

%

The normalization of the dynamical generators is
not fixed by the superalgebra at order ${\mc O}(g_2)$ and can be an arbitrary (dimensionless)
function $f(\m\a_3\,,\,\frac{\a_1}{\a_3})$ of the light-cone
momenta and $\m$ due to the fact that $P^+$ is a central element of the algebra.
Indeed, it does not seem that further consistency conditions at higher orders in $g_2$ would allow to fix $f$.

Other solutions of the dynamical constraints to this order in $g_2$ can be found
and have been provided in~\cite{DiVecchia:2003yp,Dobashi:2004nm}.

Consider now the constraints at order ${\mc O}(g_2^2)$. These are, 
\bea
&&\{Q_{2\,\a_1\da_2},Q_{4\,\b_1\db_2}\}+\{Q_{4\,\a_1\da_2},Q_{2\,\b_1\db_2}\}
+ \{Q_{3\,\a_1\da_2},Q_{3\,\b_1\db_2}\}  = -2\e_{\a_1\b_1}\e_{\da_2\db_2}H_4\,,\label{dyn4}\\
&&\{\wt{Q}_{2\,\a_1\da_2},\wt{Q}_{4\,\b_1\db_2}\}+\{\wt{Q}_{4\,\a_1\da_2},\wt{Q}_{2\,\b_1\db_2}\}
+ \{\wt{Q}_{3\,\a_1\da_2},\wt{Q}_{3\,\b_1\db_2}\} =
-2\e_{\a_1\b_1}\e_{\da_2\db_2}H_4\,,\label{dyn5}\\
&&\{Q_{2\,\a_1\da_2},\wt{Q}_{4\,\b_1\db_2}\} +
\{Q_{4\,\a_1\da_2},\wt{Q}_{2\,\b_1\db_2}\} +
\{Q_{3\,\a_1\da_2},\wt{Q}_{3\,\b_1\db_2}\}= 0,\label{dyn6}\, \eea
from these we derive the contact term used in the calculations. As it
stands $Q_4$ is unknown and the custom in the literature is to set
it to zero. Note that this choice is not inconsistent with the
constraints listed above. Setting $Q_4$ to zero, we arrive at the
following form of $H_4$ which is used in the contact term calculations
of sections \ref{trace} and \ref{4imp},

\be \label{contact}
H_4=\frac{1}{8}
Q_{3\beta_1\dot\beta_2}Q_3^{\beta_1\dot\beta_2}+\frac{1}{8}Q_{3\dot\beta_1\beta_2}
Q_3^{\dot\beta_1\beta_2}. \ee


\section{BMN basis}
\label{appA}

In the calculations we use
for the oscillators the BMN basis which is related to the exponential basis by
\bea
&&\sqrt{2}a_n^i \equiv \a_n^i+\a_{-n}^i\,,~~~i\sqrt{2}a_{-n}^i \equiv \a_n^i-\a_{-n}^i\,\cr
&&\sqrt{2}a_n^{i'}\equiv \a_n^{i'}+\a_{-n}^{i'}\,,~~~i\sqrt{2}a_{-n}^{i'} \equiv \a_n^{i'}-\a_{-n}^{i'}\,
,\cr
&&\sqrt{2}b_n^{\a_1\a_2}\equiv \b_n^{\a_1\a_2}+\b_{-n}^{\a_1\a_2}\,,~~~
i\sqrt{2}b_{-n}^{\a_1\a_2} \equiv \b_n^{\a_1\a_2}-\b_{-n}^{\a_1\a_2}\,,\cr
&&i\sqrt{2}b_n^{\da_1\da_2}\equiv - \b_n^{\da_1\da_2}+\b_{-n}^{\da_1\da_2}\,,~~~
\sqrt{2}b_{-n}^{\da_1\da_2} \equiv \b_n^{\da_1\da_2}+\b_{-n}^{\da_1\da_2}\,~~~~
\label{basis}
\eea
for $n>0$, and
\be
a_0^i \equiv \a_0^i\,~~~a_0^{i'} \equiv \a_0^{i'}\,~~~b_0^{\a_1\a_2} \equiv \b_0^{\a_1\a_2}\,
~~~\sqrt{2}b_{-n}^{\da_1\da_2} \equiv \b_n^{\da_1\da_2}+\b_{-n}^{\da_1\da_2}\,
\ee
for $n=0$.

The commutation relation for the oscillators are
\begin{equation}
[\a_m^i,\a_n^{\dag\,j}]  = \d_{mn}\d^{ij}\,, \quad \{\bigl(\b_m\bigr)_{\a_1\a_2},\bigl(\b_n^{\dag}\bigr)^{\b_1\b_2}\}
= \d_{mn}\d^{\b_1}_{\a_1}\d^{\b_2}_{\a_2}\,.
\end{equation}
The following relations are useful
\begin{equation}
[\bigl(\a_m\bigr)_{\a_1\da_1},\bigl(\a_n^{\dag}\bigr)^{\db_1\b_1}] = \d_{mn}\d^{\b_1}_{\a_1}\d^{\db_1}_{\da_1}\,,\qquad
[\bigl(\a_m\bigr)_{\a_2\da_2},\bigl(\a_n^{\dag}\bigr)^{\db_2\b_2}] = \d_{mn}\d^{\b_2}_{\a_2}\d^{\db_2}_{\da_2}\,.
\end{equation}
where
\begin{equation}
\bigl(\a_n^{\dag}\bigr)_{\a_1\da_1}  \equiv \frac{1}{\sqrt{2}}\bigl(\s^i\bigr)_{\a_1\da_1}\a_n^{\dag\,i}\,, \quad
\bigl(\a_n^{\dag}\bigr)_{\a_2\da_2}  \equiv \frac{1}{\sqrt{2}}\bigl(\s^{i'}\bigr)_{\a_2\da_2}\a_n^{\dag\,i'}
\end{equation}

The free light-cone Hamiltonian (\ref{h2ab}) becomes
\be
H_{2(r)}=\frac{1}{\alpha_r}\sum_{n\in {\mathcal Z}}\omega_{n(r)} N_{n(r)}
\label{freeh}
\ee
where
$N_{n(r)}$ is the number operator
\begin{equation}
N_{n(r)} = \a_{n(r)}^{\dag\,i}\a_{n(r)}^i+\a_{n(r)}^{\dag\,i'}\a_{n(r)}^{i'}
+\bigl(\b_{n(r)}^{\dag}\bigr)^{\a_1\a_2}\bigl(\b_{n(r)}\bigr)_{\a_1\a_2}+
\bigl(\b_{n(r)}^{\dag}\bigr)^{\da_1\da_2}\bigl(\b_{n(r)}\bigr)_{\da_1\da_2}\,.
\end{equation}
The ground state is defined as
\begin{equation}
\a_{n(r)}|\a_r\ra = 0\,,\quad \b_{n(r)}|\a_r\ra = 0\,,\quad n\in{\mathcal Z}\,.
\end{equation}
The free dynamical supercharges (\ref{q-mode}) are given by
\bea
\sqrt{\frac{|\a|}{2}}Q^-_{\a_1\da_2} &=&-\frac{\sqrt{\m|\a|}}{2\sqrt{2}}(1-e(\a))
\left[\a_{0\,\a_1}^{\,\,\,\,\db_1}\b_{0\,\db_1\da_2}^{\dag}+\a_{0\,\da_2}^{\dag\,\b_2}\b_{0\,\a_1\b_2}\right]\cr
&+&\sum_{k\neq 0}
\left[\sqrt{\omega_k+\m\a}\,\a^{\dag\,\db_1}_{k\,\a_1}\b_{k\,\db_1\da_2}-ie(\a k)\sqrt{\omega_k-\m\a}\,
\a_{k\,\a_1}^{\,\,\,\,\db_1}\b_{k\,\db_1\da_2}^{\dag}\right.\cr
&&\left.-e(\a)\left(\sqrt{\omega_k+\m\a}\,\a_{k\,\da_2}^{\b_2}\b_{k\,\a_1\b_2}^{\dag}-ie(\a k)
\sqrt{\omega_k-\m\a}\,\a_{k\,\da_2}^{\dag\,\b_2}\b_{k\,\a_1\b_2}\right)\right]\,,\cr
\sqrt{\frac{|\a|}{2}}Q^-_{\da_1\a_2}& =& \frac{\sqrt{\m|\a|}}{2\sqrt{2}}(1+e(\a))
\left[\a_{0\,\da_1}^{\,\,\,\,\b_1}\b_{0\,\b_1\a_2}^{\dag}+\a_{0\,\a_2}^{\dag\,\db_2}\b_{0\,\da_1\db_2}\right]\cr
&+&\sum_{k\neq 0}
\left[\sqrt{\omega_k+\m\a}\,\a^{\dag\,\db_2}_{k\,\a_2}\b_{k\,\da_1\db_2}-ie(\a k)\sqrt{\omega_k-\m\a}\,
\a_{k\,\a_2}^{\,\,\,\,\db_2}\b_{k\,\da_1\db_2}^{\dag}\right.\cr
&&\left.+e(\a)\left(\sqrt{\omega_k+\m\a}\,\a_{k\,\da_1}^{\,\,\,\,\b_1}\b_{k\,\b_1\a_2}^{\dag}-ie(\a k)
\sqrt{\omega_k-\m\a}\,\a_{k\,\da_1}^{\dag\,\b_1}\b_{k\,\b_1\a_2}\right)\right]\,
\eea
and $\bar{Q}^-=e(\a)\bigl[Q^-\bigr]^{\dag}$.

The calculations in sections \ref{trace} and \ref{4imp} use the
following expressions for the cubic vertex and supercharges, which can be deduced
from their forms given in the previous Appendix and 
are valid for bosonic external and intermediate states for $H_3$ and
intermediate states involving a single fermionic (and an arbitrary
number of bosonic) impurities for $H_4$

\begin{equation}\label{H3}
|H_3\rangle = -g_2 \frac{r(1-r)}{4} \sum_{s=1}^3 \sum_{p=-\infty}^{\infty}
\sum_{K,L=1}^{8} \frac{\omega_{p(s)}}{\alpha_s} \alpha_{p\,(s)}^{\dagger K} \alpha_{-p\,(s)}^{L}
\,\Pi^{KL}
|V\rangle
\end{equation}

\begin{equation}\label{Q}
|Q_{3\, \beta_1 {\dot \beta}_2} \rangle = g_2 \frac{\eta}{4}
\sqrt{\frac{r\,(1-r)\, \alpha'}{-2\,\alpha_3^3}}
\sum_{r,s=1}^3 \sum_{p,q=-\infty}^{\infty}
(\sigma^k)^{{\dot \gamma}_1}_{\beta_1}  K_{p\,(s)} G_{|q|(r)} \,
\alpha_{-p\,(s)}^{\dagger k} \beta^{\dagger}_{q\,(r)\, {\dot \gamma}_1 {\dot \beta}_2}
|V\rangle
\end{equation}

\begin{equation}\label{Qbar}
|Q_{3\, {\dot \beta}_1 \beta_2} \rangle = g_2 \frac{{\bar \eta}}{4}
\sqrt{\frac{r\,(1-r)\, \alpha'}{-2\,\alpha_3^3}}
\sum_{r,s=1}^3 \sum_{p,q=-\infty}^{\infty}
(\sigma^k)^{\gamma_1}_{{\dot \beta}_1}  K_{p\,(s)} G_{|q|(r)} \,
\alpha_{-p\,(s)}^{\dagger k} \beta^{\dagger}_{q\,(r)\, \gamma_1 \beta_2}
|V\rangle
\end{equation}
and similarly for $|\widetilde{Q}_3\rangle$.
The kinematical part of the vertex $|V\rangle$ in the number basis is defined as follows,

\begin{equation}
|V \rangle = |E_{\alpha}\rangle |E_{\beta}\rangle\delta(\sum_{r=1}^3 \alpha_r)
\ee
where $|E_{\alpha}\rangle$ and $|E_{\beta}\rangle$ are exponentials of bosonic and fermionic
oscillators respectively
\be
|E_{\alpha}\rangle=
\exp \left( \frac{1}{2} \sum_{r,s=1}^{3} \sum_{m,n = -\infty}^{\infty}
\alpha^{\dagger K}_{m\,(s)} {\widetilde N}^{st}_{mn} \alpha^{\dagger K}_{n\,(t)} \right )
| \alpha \rangle _{123}
\label{eb}
\end{equation}
and
\begin{equation}
|E_{\beta}\rangle=\exp\left(
\sum_{r,s=1}^3\sum_{m,n= -\infty}^{\infty}
\bigl(\b^{\a_1\a_2\,\dag}_{m(r)}\b^{\dag}_{n(s)\,\a_1\a_2}-
\b^{\da_1\da_2\,\dag}_{m(r)}\b^{\dag}_{n(s)\,\da_1\da_2}\bigr)
\wt{Q}^{rs}_{mn}\right)| \alpha \rangle _{123}
\label{ef}
\end{equation}
$| \alpha \rangle _{123}=|\alpha_1 \rangle \otimes|\alpha_2 \rangle\otimes|\alpha_3 \rangle$
is the tensor product of three vacuum states.
All the quantities appearing above are defined in Appendix
\ref{neu}.

\section{Neumann matrices and associated quantities}
\label{neu}
In this section we present the explicit expressions for the quantities appearing in the prefactors and exponentials part
of $|H_3\rangle$ and $|Q_3\rangle$.
Following the notation of \cite{Gutjahr:2004dv}, the Neumann matrices can be written as

\bea
&&\wt{N}^{st}_{mn}=
\cases{
\frac{1}{2}\bar{N}^{st}_{|m||n|}\left(1+U_{m(s)}U_{n(t)}\right)~~~~,m,n\neq0 \cr
\frac{1}{\sqrt{2}}\bar{N}^{st}_{|m|0}~~~~ m\neq 0 \cr
\bar{N}^{st}_{00}\,}
\eea
with\footnote{To have a manifest symmetry in $1\leftrightarrow 2$ we additionally redefined the oscillators
as $(-1)^{s(n+1)}\a_{n(s)} \to \a_{n(s)}$ for
$n\in{\mathcal Z}$, $s=1,2,3$ and analogously for the fermionic oscillators.}
\be
\label{mn}
\bar{N}^{st}_{mn} =-(1-4\m\k K)^{-1}\frac{\k}{\a_s\omega_{n(t)}+
\a_t\omega_{m(s)}}\left[CU_{(s)}^{-1}C_{(s)}^{1/2}\bar{N}^s\right]_m
\left[CU_{(t)}^{-1}C_{(t)}^{1/2}\bar{N}^t\right]_n\,
\ee
\be
\label{m0}
\bar{N}^{st}_{m0} =
\sqrt{-2\m\k(1-\b_t)}\sqrt{\omega_{m(s)}}\bar{N}^s_m\,,\qquad t\in\{1,2\}
\ee
\be
\label{00a}
\bar{N}^{st}_{00} =
(1-4\m\k K)\left(\d^{st}-\sqrt{\b_s\b_t}\right)\,,\qquad  s,t\in\{1,2\}\,
\ee
\be
\label{00b} \bar{N}^{s3}_{00} = -\sqrt{\b_s}\,,\qquad s\in\{1,2\}
\ee
where
\be
C_n = n\,,\qquad C_{n(s)} = \omega_{n(s)}\equiv\sqrt{n^2+\bigl(\m\a_s\bigr)^2}\,,\qquad \k\equiv \a_1\a_2\a_3
\ee
\be
U_{n(s)} =\frac{1}{n}(\omega_{n(s)}-\m\a_s)\,,\qquad U^{-1}_{n(s)}=\frac{1}{n}(\omega_{n(s)}+\m\a_s)
\ee
and~\cite{He:2002zu}
\be
\label{K}
1-4\m\k K \approx -\frac{1}{4\pi r(1-r)\m\a_3}\,
\ee
\be
\label{N3}
\a_3\bar{N}^3_n \approx -\frac{\sin(n\pi r)}{\pi r(1-r)}\frac{1}{\omega_{n(3)}
\sqrt{-2\m\a_3(\omega_{n(3)}+\m\a_3)}}\,
\ee
\be
\label{Nr}
\a_3\bar{N}^s_n \equiv \a_3\bar{N}_n(\b_s) \approx -\frac{\sqrt{\b_s}}{2\pi r(1-r)}\frac{1}
{\omega_{n(s)}\sqrt{-2\m\a_3(\omega_{n(s)}-\m\a_3\b_s)}}
\ee
up to exponential corrections $\sim{\mc O}(e^{-\m\a_3})$
\footnote{To compare with the definition used in~\cite{He:2002zu} note that
$\bar{N}^s_{n\,\mbox{here}}=(-1)^{s(n+1)}U_{n(s)}C_{n(s)}^{-1/2}\bar{N}^s_{n\,\mbox{there}}$.}.
For the bosonic constituents of the prefactor one has
\be
\label{k}
K^I = \sum_{s=1}^3\sum_{n\in{\mathcal Z}}K_{n(s)}\a_{n(s)}^{I\,\dagger}\,,~~~
\wt{K}^I= \sum_{s=1}^3\sum_{n\in{\mathcal Z}}K_{n(s)}\a_{-n(s)}^{I\,\dagger}\,
\ee
where
\be
K_{0(s)}= (1-4\m\k K)^{1/2}\sqrt{-\frac{2\m\k}{\a'}\bigl(1-\b_s\bigr)}\,,\qquad K_{0(3)} = 0\,
\ee
and
\be
K_{n(s)}= -\frac{\k}{\sqrt{2\a'}\a_s}(1-4\m\k K)^{-1/2}(\omega_{n(s)}+\m\a_s)
\sqrt{\omega_{n(s)}}\bar{N}^s_{|n|}\bigl(1-U_{n(s)}\bigr)\,
\ee
For the fermionic constituents of the prefactor one has
\be
\label{yz}
Y^{\a_1\a_2} = \sum_{s=1}^3\sum_{n\in{\mathcal Z}}G_{|n|(s)}\b^{\dag\,\a_1\a_2}_{n(s)}\,,\qquad
Z^{\da_1\da_2} = \sum_{s=1}^3\sum_{n\in{\mathcal Z}}G_{|n|(s)}\b^{\dag\,\da_1\da_2}_{n(s)}\,,
\ee
where
\be
G_{0(s)} = (1-4\m\k K)^{1/2}\sqrt{1-\b_s}\,,\qquad G_{0(3)} = 0\,
\ee
and
\be
G_{n(s)}= \frac{e(\a_s)}{\sqrt{2|\a_s|}}\frac{\sqrt{-\k}}{(1-4\m\k K)^{1/2}}
\sqrt{(\omega_{n(s)}+\m\a_s)\omega_{n(s)}}\bar{N}^s_{|n|}\,
\ee
where in the above expressions we have used $\b_1\equiv r$ and $\b_2 \equiv 1-r$
(with $\b_t\equiv-\a_t/\a_3$ and $\a_3 < 0$).

Let us now collect some expressions needed for the computations presented in the Paper.
The Neumann matrices which couple the external string
(labeled by 3) to the internal strings which are labeled by $r,s =
1,2$ are special in that they contain a pole proportional to the
external state mode number. Taking a large $\mu$ limit of the
expression given above we get,

\begin{equation}
{\widetilde N}^{3\,r}_{n\, p} \simeq e(n) \, \frac{ \sin(|n|\pi r) }
{2 \pi \sqrt{\omega^{(3)}_n \, \omega^{(r)}_p } }
\, \frac{ \left( \omega^{(r)}_p + \beta_r \, \omega^{(3)}_n \right) }{p - \beta_r\,n }
\end{equation}
where $\beta_1=r$, $\beta_2=(1-r)$ ($\beta_r=-\alpha_r/\alpha_3$).

The internal-internal Neumann matrix is,

\begin{equation}
{\widetilde N}^{r\,s}_{n\, p} = 
\frac{ \sqrt{\beta_r\,\beta_s} \left(
\sqrt{ \omega^{(r)}_n - \beta_r \mu \alpha_3  }
\sqrt{ \omega^{(s)}_p - \beta_s \mu \alpha_3  }
+e(n\, p)\,
\sqrt{ \omega^{(r)}_n + \beta_r \mu \alpha_3  }
\sqrt{ \omega^{(s)}_p + \beta_s \mu \alpha_3  }
\right) }{4\pi \sqrt{\omega^{(r)}_n \omega^{(s)}_p  }
\left( \beta_s \omega^{(r)}_n + \beta_r  \omega^{(s)}_p \right) }
\end{equation}

 The $K$ and $G$ vectors from the prefactor are as follows,

\begin{equation}
K^{(r)}_{-p} = -\alpha_3 \,\sqrt{\frac{r\,(1-r)}{4\pi\,\beta_r\,\alpha'}}
\,\frac{ \sqrt{ \omega^{(r)}_p - \beta_r \mu \alpha_3 }
+ e(p) \sqrt{ \omega^{(r)}_p +\beta_r \mu \alpha_3 } }
{\sqrt{ \omega^{(r)}_p} }
\end{equation}

\begin{equation}
K^{(3)}_{-n} = -\alpha_3 \,\sin(|n|\pi r) \sqrt{\frac{r\,(1-r)}{\pi\,\alpha'}}
\,\frac{ \sqrt{ \omega^{(3)}_n + \mu \alpha_3 }
+ e(n) \sqrt{ \omega^{(3)}_n -\mu \alpha_3 } }
{\sqrt{ \omega^{(3)}_n} }
\end{equation}

\begin{equation}
G^{(r)}_p = \frac{1}{\sqrt{4\pi\, \omega^{(r)}_p } } \qquad
G^{(3)}_n = -\frac{ \sin(|n|\pi r) }{\sqrt{\pi\,\omega^{(3)}_n}}
\end{equation}

 Finally, the fermionic Neumann matrices are given by,

\begin{equation}
{\widetilde Q}^{r\,s}_{n\, p} = i\,\frac{\beta_s}{4\pi}
\frac{n}{ \sqrt{\omega^{(r)}_n \omega^{(s)}_p  }
\left( \beta_s \omega^{(r)}_n + \beta_r  \omega^{(s)}_p \right) }
\end{equation}

\begin{equation}
{\widetilde Q}^{3\,r}_{n\, p} - {\widetilde Q}^{r\,3}_{p\, n} = -i
\frac{ \sin(|n|\pi r) }
{2 \pi \sqrt{\omega^{(3)}_n \, \omega^{(r)}_p } }
\, \frac{ \left( \omega^{(r)}_p + \beta_r \, \omega^{(3)}_n \right) }{p - \beta_r\,n }
\end{equation}

\section{Leading $\mu$ dependence of sums}
\label{largemu}

\indent Sums are evaluated using the contour integral method,

\begin{equation}\label{contmeth}
\sum_{p=-\infty}^{\infty} f(p) = -\frac{i}{2}\oint dz \, f(z) \, \cot(\pi z)
\end{equation}
rotating and scaling the integration variable through the
substitution $z \rightarrow i x z$, where $x = -\mu\alpha_3$,
turns the cotangent into $\coth(\pi x z)$ which can be set to one
in the large $x$ limit. If the summand $f(z)$ has no poles on the
real axis, the procedure simply replaces $p$ by $p' = x\,p$ and
integrates,

\begin{equation}\label{nopoles}
\sum_{p=-\infty}^{\infty} f(p) = \int_{-\infty}^\infty dp' f(p')
\end{equation}
yielding the large $x$ behavior. If there are poles on the real axis,
one must evaluate their residue using the integrand in (\ref{contmeth})
and then integrate along any cut which $f(z)$ may possess along the
imaginary axis. The essential point here is that the cut integrals
are always sub-leading compared to the residues, so that the sum is dominated
by the summand's behavior at the poles. For the sums which concern us
in this paper, poles come from factors of ${\widetilde N}^{3\,r}_{n\, p}$
which as far as power counting in $x$ is concerned should simply be ignored,
replacing the summation variable $p$ with it's value at the pole everywhere
in the summand and adding some factors of $\pi$ and $\cot(\pi n)$.
By contrast sums not involving ${\widetilde N}^{3\,r}_{n\, p}$ can be evaluated
by straightforward application of (\ref{nopoles}).

\end{document}